\documentclass[11pt]{article}

\usepackage{amsmath,amssymb,graphicx}

\def\RR{\mathbb R}
\def\ZZ{\mathbb Z}

\def\cA{\mathcal A}

\def\cC{\mathcal C}
\def\cD{\mathcal D}
\def\cF{\mathcal F}
\def\cG{\mathcal G}

\def\cH{\mathcal H}

\def\cP{\mathcal P}
\def\cL{\mathcal L}
\def\cT{\mathcal T}

\def\cH{\mathcal H}

\def\cX{\mathcal X}

\def\cU{\mathcal U}
\def\cV{\mathcal V}

\def\bzero{\mathbf 0}

\newcommand{\hide}[1]{}
\newcommand{\raf}[1]{(\ref{#1})}

\newcommand{\Tr}{\operatorname{Tr}}
\newcommand{\Min}{\operatorname{Min}}
\newcommand{\poly}{\operatorname{poly}}

\def\T{T}
\newcommand{\proof}{\noindent {\bf Proof}.~~}
\newcommand{\qed}{\hfill$\square$\bigskip}

\newcommand{\VD}[2]{Dec$(#1;\cX\subseteq #2)$}

\newtheorem{theorem}{Theorem}
\newtheorem{fact}{Fact}
\newtheorem{lemma}{Lemma}
\newtheorem{corollary}{Corollary}
\newtheorem{proposition}{Proposition}

\newtheorem{remark}{Remark}

\title{Enumerating Vertices of  $0/1$-Polyhedra associated with $0/1$-Totally Unimodular Matrices}
\author{
Khaled Elbassioni\thanks{Masdar Institute, Khalifa University of Science and Technology, P.O. Box 54224, Abu Dhabi, UAE;
(kelbassioni@masdar.ac.ae)}
\and
Kazuhisa Makino\thanks{Research Institute for Mathematical Sciences (RIMS)
Kyoto University, Kyoto 606-8502, Japan;
(makino@kurims.kyoto-u.ac.jp)}
}

\begin{document}
\date{}
\maketitle

\begin{abstract}
	We give an incremental polynomial time algorithm for enumerating the vertices of any polyhedron $\cP(A,\b1)=\{x\in\RR^n \mid Ax\geq \b1,~x\geq \b0\}$, when $A$ is a totally unimodular matrix. Our algorithm is based on decomposing the hypergraph transversal problem for unimodular hypergraphs using Seymour's decomposition of totally unimodular matrices, and  may be of independent interest.
	% and its extension to $0/1$-totally unimodular matrices by Truemper.

\end{abstract}

\section{Introduction}\label{s-intro}

\subsection{The vertex enumeraion problem}
The well-known Minkowski-Weyl theorem states that any convex polyhedron $\cP\subseteq\RR^n$
can be represented as the Minkowski sum of the convex hull
of the set $\cV(\cP)$ of its extreme points and the conic hull of the set $\cD(\cP)$
of its extreme directions (see e.g. \cite{S86}). 
%Furthermore, for pointed polyhedra, i.e., those that do not contain lines, this representation is unique.
Given a polyhedron $\cP$ by its linear description %(called $\cH$-representation)
as the intersection of finitely many halfspaces, obtaining the set $\cV(\cP)\cup\cD(\cP)$,
required by the other representation, %(called $\cV$-representation), 
is a well-known problem, called \emph{Vertex Enumeration} (VE) (see,. e.g.,\cite{Dye83,BFM98}), which have been extensively
studied in the literature in different (but polynomially equivalent) forms, e.g.,
, the \emph{facet enumeration} problem \cite{BFM98}
or the \emph{polytope-polyhedron problem} \cite{Lov92}. 
%In the following, we will write $\cP_\cH$ and $\cP_\cV$ to distinguish whether
%$\cP$ is given by its $\cH$-representation or $\cV$-representation, respectively.
Clearly, the size of the extreme set $\cV(\cP)\cup\cD(\cP)$ can be (and typically is)
exponential in the dimension $n$ and the number of linear inequalities $m$, and thus when
considering the computational complexity of the vertex enumeration problem,
one is usually interested in \emph{output-sensitive} algorithms \cite{Sei86}, i.e.,
those whose running time depends not only on $n$ and $m$, but also on $|\cV(\cP)\cup\cD(\cP)|$.
Alternatively, we may consider the following, polynomially equivalent, decision variant of the problem:

\begin{description}
	\item [\VD{\cL}{\cC(\cP)}:] Given a polyhedron $\cP$, represented by a system of linear inequalities $\cL$,
	and a subset $\cX\subseteq\cC(\cP)$, is $\cX=\cC(\cP)$?
	%\item[\VDD{P}:] Given a polyhedron $P$, represented
	%by a system of linear inequalities, and a subset of its
	%extreme  set $\cX\subseteq\cV(P)\cup\cD(P)$, is $\cX=\cV(P)\cup\cD(P)$?.
\end{description} 
In this description, $\cC(\cP)$ could be either $\cV(\cP)$, $\cD(\cP)$, or $\cV(P)\cup\cD(P)$. 
%It is well-known and also easy to see that the decision problems for $\cD(\cP)$ or for $\cV(\cP)\cup\cD(\cP)$
%are equivalent to that for $\cV(\cP')$ where $\cP'$ is some polytope
%whose description size is polynomial in that of $\cP$.  
%% by shifting all facet-defining hyperplanes to the origin, and adding a normalization constraint.   
The problem of enumerating the elements of $\cC(\cP)$ is said to be solvable in {\it incremental polynomial} time if problem \VD{\cL}{\cC(\cP)} can be solved in time polynomial in the size of the description of $\cL$ and $\cX$.\footnote{Note that if the answer to the decision problem is ``NO" then a new element in $\cC(\cP)\setminus\cX$ can be found by a polynomial number of calls to the decision problem.}    
It is 
%also 
well-known that if the decision problem is NP-hard, then no {\it output} (or {\it total}) {\it polynomial-time} algorithm
can generate the elements of $\cC(\cP)$ unless P=NP (see e.g. \cite{BEGM09}). 

Vertex enumeration is an outstanding open problem in computational geometry and polyhedral combinatorics
(see, e.g., \cite{DP77,Lov92,Pfe02}), and has numerous
applications. For example, understanding the structure 
of the vertices helps in designing approximation algorithms for
combinatorial optimization problems \cite{V01}; finding all vertices
can be used for computing Nash equilibria for bimatrix games
\cite{ARSS10}.        
Numerous algorithmic ideas for vertex or facet enumeration have been introduced in the literature, see, e.g.,   
\cite{Abd03,ABS97,AF92,AF96,BFM98,BL98,Chv83,DP77,Dye83,FP96,Pro94,RS14,Sei86}.

\medskip

The main result in \cite{KBBEG08} established that problem
\VD{\cL}{\cV(\cP)} is NP-hard for unbounded polyhedra,
more precisely, when $|\cD(\cP)|$ is exponentially large in the input 
size. This negative result holds, even when restricted to $0/1$-{\it
	polyhedra}  \cite{BEGT11}, that is, when
$\cV(\cP)\subseteq\{0,1\}^n$, and comes in contrast with the fact that
 the VE problem for $0/1$-polytopes (i.e., bounded polyhedra) is known to be solvable with
{\it polynomial delay} (that is, the vertices are generated such that the
delay between any successive outputs is polynomial {\it only} in the {\it input}
size) and {\it polynomial space} (that is, the total space used for enumerating all the vertices is polynomial in the {\it input}
size).

\subsection{VE for $0/1$-polyhedra associated with $0/1$-totally unimodular matrices}

\medskip
Let $A\in\{0,1\}^{m\times n}$ be an $m \times n$ $0/1$-matrix such that the
polyhedron 
\begin{equation}\label{P}
\cP(A,\b1)=\{x\in\RR^n \mid Ax\geq \b1,~x\geq \b0\}
\end{equation}
has only integral vertices, where $\b1$ (resp., $\b0$) denotes the vector of all ones (resp., zeros) of appropriate dimension.
Then $\cP(A,\b1)$ has only $n$ extreme directions (namely the $n$ unit vectors in $\RR^n$), while the vertices of $\cP(A,\b1)$ are in one-to-one correspondence with the minimal transversals of the hypergraph $\cH[A] \subseteq  2^{[n]}$, whose characteristic vectors of hyperedges are the rows of $A$. One of the most important examples is when the matrix $A$ is \emph{totally unimodular}: in this case, the polyhedron $\cP(A,\b1)$ has integral vertices, and VE is equivalent to finding all minimal transversals\footnote{Note that, it is not possible to reduce the problem of enumerating the vertices of $\cP(A,\b1)$ to that of enumerating the vertices of the $0/1$ polytope $\cP'=\{x\in\RR^n \mid Ax\geq \b1,~\b0\leq x\leq\b1\}$, as $\cP'$ can have exponentially more vertcies than those of $\cP$ (namely, the vertices of $\cP'$ are the (not necessarily minimal) transversals of $\cH[A]$).} of a \emph{unimodular hypergraph} $\cH[A]$. Consequently, it follows from the well-known result in \cite{FK96} that all vetrices of such polyhedra can be enumerated in {\it quasi-polynomial} time, and hence the VE problem in this case is unlikely to be NP-hard.  Polynomial time algorithms for special cases of this problem  are known; for example, enumerating minimal vertex/edge covers in bipartite graphs \cite{EG95,MP97}, enumerating minimal hitting sets/set covers of interval hypergraphs~\cite{BEGM09}, and enumerating minimal path covers/cut conjunctions in directed trees \cite{BEGM09}. However, the complexity of the VE problem for \raf{P} remains open, even for the totally unimodular matrices $A$. In this paper, we settle the complexity of the VE problem in the latter case. 

\begin{theorem}\label{t-main}
	Let $A\in\{0,1\}^{m\times n}$ be a totally unimodular matrix. Then the vertices of $\cP(A,\b1)$ can be enumerated in incremental polynomial time.
\end{theorem} 

A celebrated result of Seymour \cite{S80} shows that any totally unimodular matrix (with $0$, $\pm1$-entries) arises from (essentially) the so-called {\it network matrices}, by a small set of simple operations. 
Similar results for $0/1$-totally unimodular matrices are derived in \cite[Chapter 11]{T92}, with the main building blocks replaced by $0/1$-network matrices.  
On the other hand, it has been shown in \cite{BEGM09} that for any polyhedron $\cP(A,\b1)$, with a $0/1$-network matrix $A$  or its transpose, the VE problem can be solved in incremental polynomial time. To prove Theorem \ref{t-main}, we show that the above mentioned decomposition of totally unimodular matrices yields a corresponding decomposition for the hypergraph transversal problem, that can be leveraged into a polynomial time algorithm for the enumeration problem.  One of the natural ways to use such decomposition is to recursively partition the input polyhedron into two smaller polyhedra and then combine the outputs from the two subproblems. While such approach works for the simple cases of the decomposition (so-called $1$- and $2$-sum decompositions), it does not work for the more complicated case (so-called $3$-sum decomposition). The main reason is that the number of vertices of the resulting polyhedra in either of the two subproblems may be exponentially larger than that in the original problem. To overcome this difficulty, we need to use the decomposition in a more sophisticated way, utilizing structural properties of the unimodular hypergraph $\cH[A]$. One technical hurdle which arises is that the total input/output size of the resulting subproblems might exceed the input/output size of the original problem, which may eventually lead to an exponential blow-up in the overall running time of the algorithm in terms of the input {\it and} output sizes. To deal with this issue, we introduce a {\it volume} measure as the product of the input and output sizes, and show in each case of our decomposition that the total measure of the  subproblems obtained is smaller than the measure of the original problem.

\section{Notation and preliminaries}\label{prelim}

\subsection{Hypergraphs and transversals}
 Let $V$ be a finite set. A hypergraph $\cH \subseteq 2^V$ is a family of subsets of $V$. A hypergraph is called {\it Sperner}  ({\it simple} or a {\it clutter}), if it has the property that no hyperedge contains another.  %$\cH$ is said to be {\it connected} if the bipartite graph $(V\cup\cH, \{(v,H)\in V\times\cH=v\in H\})$ is connected.   
For a hypergraph $\cH\subseteq 2^V$, we denote by $\Tr(\cH)$ the family of {\it minimal transversals} of $\cH$, i.e.,  (inclusion-wise) minimal subsets of $V$ which have a nonempty intersection with each hyperedge of $\cH$; $\Tr(\cH)$ is also called the {\it dual} of $\cH$. 
%For simplicity we will (sometimes) write $\Tr(\cH[A])$ as $\Tr[A]$ and also say that a set $T$ is a (minimal) transversal of $A$ to mean a (minimal) transversal of $\cH[A]$ . 
We say that the hypergraph $\cH$ is {\it trivial} if $\cH=\emptyset$ or $\cH=\{\emptyset\}$, and is {\it irredundant} if every $v\in V$ belongs to some $H\in\cH$. 
As usual, we assume $\Tr(\{\emptyset\})=\emptyset$ and $\Tr(\emptyset)=\{\emptyset\}$.

Given two hypergraphs $\cH_1$ and $\cH_2$ with vertex set $V$, denote by
\begin{align*}
 \cH_1 \wedge \cH_2 &= \Min \{ H_1 \cup H_2~ |~ H_1 \in \cH_1 \text{ and } H_2 \in \cH_2 \},\\
 \cH_1 \vee \cH_2 &= \Min (\cH_1 \cup \cH_2),
\end{align*}
the {\it conjunction} and {\it disjunction} of $\cH_1$ and $\cH_2$ respectively, where for hypergraph $\cH$, $\Min(\cH)$ denotes the family of (inclusion-wise) minimal sets in $\cH$. We denote by $\cH_1\dot{\cup}\cH_2$ the {\it disjoint} union of $\cH_1$ and $\cH_2$. For two hypergraphs $\cH_1\subseteq 2^{V_1}$ and $\cH_2\subseteq 2^{V_2}$, we denote by $\cH_1\dot{\wedge}\cH_2$ the conjunction of $\cH_1$ and $\cH_2$ when $V_1$ and $V_2$ are {\it disjoint}. By definition, $|\cH_1\dot{\cup}\cH_2|=|\cH_1|+|\cH_2|$ and $|\cH_1\dot{\wedge}\cH_2|=|\cH_1|\cdot|\cH_2|$.

For a hypergraph $\cH\subseteq 2^V$ and a set $S\subseteq V$, we denote by $\cH_S=\{H\in\cH~ |~ H\subseteq S\}$ and $\cH^S=\Min\{H\cap S~ |~ H\in\cH\}$ 
 the subhypergraph of $\cH$ {\it induced by} $S$, and the {\it projection} of $\cH$ on $S$, respectively. For $W,S\subseteq V$, we write $\cH(W,S)=\{H\in\cH\mid H\cap W=S\}$. Two vertices of $\cH$ are said to be {\it identical} if they belong to exactly the same hyperedges, i.e., the corresponding columns in the hyperedge-vertex incidence matrix are identical.

The following propositions are straightforward (see e.g. \cite{B89,EGM03,LLK80}).
\begin{proposition}\label{prop1}
Given a hypergraph $\cH\subseteq 2^V$ and a set $S\subseteq V$, the following statements hold:
\begin{itemize}
\item[(i)]   $\Tr(\Tr(\cH))=\Min(\cH)$,
\item[(ii)]  $ \Tr(\cH_S) = \Tr(\cH)^S $ (and hence, $ \Tr(\cH^S) = \Tr(\cH)_S $) and
\item[(iii)] $ |\Tr(\cH_S)| \leq |\Tr(\cH)| $.
\end{itemize}
\end{proposition}
\begin{proposition}\label{prop2}
Given hypergraphs $\cH_1,\ldots,\cH_k \subseteq 2^V$,
\[ \Tr\left(\bigvee_{i=1}^r\cH_i\right)= \bigwedge_{i=1}^r\Tr(\cH_i). \]
\end{proposition}
As a corollary of Proposition~\ref{prop2} we have the following.
\begin{proposition}\label{p2}
	Let $\cH\subseteq 2^V$ be a hypergraph and $S_1,\ldots,S_r\subseteq V$
	be subsets such that for every hyperhedge $H\in\cH$ there exists an
	$i\in\{1,\ldots,r\}$ with $H\subseteq S_i$. Then
	\begin{equation*}\label{conj}
	\Tr(\cH)=\bigwedge_{i=1}^r\Tr(\cH_{S_i}).
	\end{equation*}
\end{proposition}

%\begin{proposition}\label{p3-}
%Let $\cH\subseteq 2^V$ be a hypergraph and $S\subseteq V$. Then $|\Tr(\cH_S)|\le|\Tr(\cH)|$.
%\end{proposition}

Throughout the paper, we use the notation: $n=n(\cH)=|V|$, $m=m(\cH)=|\cH|$ and   $k=k(\cH)=|\Tr(\cH)|$.

\subsection{Polyhedra}

A convex polyhedron $P\subseteq\RR^n$ is the intersection of finitely many halfspaces, determined by the \emph{facets} of the polyhedron.
A \emph{vertex} or an \emph{extreme point} of $P$ is a point $v\in\RR^n$ which cannot be represented as a convex combination of two other points of $P$, i.e., there exists no $\lambda\in(0,1)$ and $v_1,v_2\in P$ such that $v=\lambda v_1+(1-\lambda)v_2$. A \emph{(recession) direction} of $P$ is a vector $d\in\RR^n$ such that $x_0+\mu d\in P$ whenever $x_0\in P$ and $\mu\ge 0$. An \emph{extreme direction} of $P$ is a direction $d$ that cannot be written as a conic combination of two other directions, i.e., 
there exist no positive real numbers $\mu_1,\mu_2\in\RR_+$ and directions $d_1,d_2$ of $P$ such that $d=\mu_1 d_1+\mu_2 d_2$.   
Denote respectively by $\cV(P)$ and $\cD(P)$ the sets of extreme points and extreme directions of polyhedron $P$. 
A bounded polyhedron, i.e., one for which $\cD(P)=\emptyset$ is called a \emph{polytope}.

\subsection{Totally unimodular matrices}
A matrix $A\in\{0,1\}^{m\times n}$ is {\it totally unimodular} if every square subdeterminant of it has value in $\{-1,0,1\}$.
We denote by $\cU^{m\times n}$ the set of $m\times n$ $0/1$-totally unimodular matrices. For a matrix $A\in\{0,1\}^{m\times n}$ we denote by $\cH[A] \subseteq  2^{[n]}$ the hypergraph whose characteristic vectors of hyperedges are the rows of $A$. %Such a matrix is called a  clutter if the corresponding hypergraph $\cH[A]$ is a clutter, and is connected if $\cH[A]$ is connected.  
A hypergraph $\cH$ is said to be {\it unimodular} \cite{B89} if $\cH=\cH[A]$ for a totally unimodular matrix $A$.  Note by definition that if $\cH\subseteq 2^V$ is unimodular then for any set $S\subseteq V$ and any subhypergraph  $\cH'\subseteq\cH$, the hypergraph $(\cH')^S$ is unimodular.
A $0/1$ matrix is said to be {\it ideal} (see, e.g., \cite{C01}) if the
polyhedron $P=P(A,\b1)$ has only integral vertices. It is well-known that every totally unimodular matrix $A\in\{0,1\}^{m\times n}$ is ideal. Furthermore, the following correspondence holds.
\begin{proposition}[\cite{L65}]\label{L}
Let $A$ be an $m \times n$ ideal matrix. 
Then the vertices of  the polyhedron $\cP(A,\b1)$ are in one-to-one correspondence with the minimal transversals of the hypergraph $\cH[A]$. 
\end{proposition}

As a corollary of Theorem~\ref{t-main}, we obtain the following result.

\begin{corollary}\label{cor1}
	Let $A\in\{0,1\}^{m\times n}$ be a totally unimodular matrix and, $A'$ be a $0/1$ matrix whose rows are the characteristic vectors of the vertices of $\cP(A,\b1)$. Then the vertices of $\cP(A',\b1)$ can be enumerated in incremental polynomial time.
\end{corollary}
\proof
By a result of Lehman \cite{L65}, the polyhedron $\cP(A',\b1)$ also is $0/1$, whenever $A'$ is a $0/1$ matrix whose rows are the characteristic vectors of the vertices of $\cP(A,\b1)$ for an deal matrix $A$.  It follows that the vertices of $\cP(A',\b1)$ are in one-to-one correspondence with the minimal transversals of $\cH[A']=\Tr(\cH[A])$.  By the polynomial equivalence of the enumeration and decision problems~\cite{BI95,GK99}, it is enough to check, for a given sublist $\cX\subseteq\Tr(\cH[A'])$, whether $\cX=\Tr(\cH[A'])$. The latter condition is equivalent to   $\Tr(\cX)=\cH[A']$, which can be checked in polynomial time by Theorem~\ref{t-main}, since $\cX$ is a unimodular hypergrpah by assumption. 
\qed

%For example, given a hypergraph $\cF\subseteq 2^V$, Corollary~\ref{cor1} implies that we can check in polynomial time in $|V|$,  $|\cF|$ and $k$ if there is a unimodular hypergraph of size at most $k$ such that $\cF=\Tr(\cH)$.  
%%%%%%%%%%%%%%%%% %%%%%%%%%%%%%%%%%%%%%%%%%%%%%%%%%%%%%%%

\subsection{$0/1$-Network matrices}
A matrix $A\in\cU^{m\times n}$ is said to be a {\it network} matrix if 
there exists a directed tree\footnote{We say that  a directed graph $G$ is a {\em directed tree} if  
 the underlying graph of $G$ (i.e., the undirected graph obtained from $G$ by ignoring
orientation of arcs) is a tree.} $\T$ such that the
rows of $A$ one-to-one correspond to the arcs in $\T$, and each 
column of $A$ is  the characteristic vector of a directed path in
$\T$.  Checking if a given matrix $A$ is a network matrix and finding the corresponding tree representation can be done in polynomial
time (see e.g., \cite{S86}). We call a hypergraph $\cH$ a {\it network hypergraph} if $\cH=\cH[A]$ for some network matrix $A$ or its transpose. It is known that network hypergraphs can be dualized in incremental polynomial time and polynomial space:

\begin{theorem}[\cite{BEGM09}]\label{t-network}
	Let $A\in\{0,1\}^{m\times n}$ be a network matrix. Then 
	\begin{description}
		\item[\hspace*{.21cm}{\rm (i)}] all the
		vertices of $\cP(A,\b1)$ can be enumerated in incremental polynomial time
		using polynomial space;
		\item[\hspace*{.21cm}{\rm (ii)}] all the vertices of
		$\cP(A^T,\b1)$ can be enumerated in incremental polynomial time
		using polynomial space.
	\end{description}
\end{theorem}

\subsection{Decomposition of $0/1$-totally unimodular matrices}
Seymour \cite{S80} gave a decomposition theorem that allows one to decompose (in polynomial time) any $0/1$-totally unimodular matrix by repeatedly applying certain operations (called {\it $i$-sums}, for $i=1,2,3$) until simple building blocks are obtained; the building blocks consist of $0/1$-network matrices, their transposes and a specific $5\times 5$ $0/1$-matrix.  
For our purposes this theorem can be stated as follows.
%This theorem has been adopted for the $0/1$-totally unimodular case by Truemper  \cite{T92} and will be summarized below.

\begin{theorem}[\cite{S80,S86,T92}]\label{t-decomp}
	Let $A\in\{0,1\}^{m\times n}$ be a totally unimodular matrix. Then one of the following conditions holds:
	\begin{itemize}
		\item[(I)] $A$ or its transpose is a $0/1$-network matrix;
		\item [(II)] $A$, possibly after permuting the rows and columns, is the matrix:  
		\begin{equation*}
		A_0=\left[\begin{array}{lllll}
		1 & 0 & 0 & 1 & 1\\
		1 & 1 & 0 & 0 & 1\\
		0 & 1 & 1 & 0 & 1\\
		0 & 0 & 1 & 1 & 1\\
		1 & 1 & 1 & 1 & 1
		\end{array}\right];
		\end{equation*}
	\item[(III)] $A$ has a row or a column with at most one ``{\rm 1"}, or  $A$ has two identical rows or columns; 	
	\item [(IV)] $A$ can be decomposed as follows:  
	\begin{equation*}
	A=\left[\begin{array}{c|c}
	A^1 & Q\\
	\hline
	R & A^2
	\end{array}\right],
	\end{equation*}
	where $A^1\in\cU^{m_1\times n_1}$, $A^2\in\cU^{m_2\times n_2},$ $Q\in\cU^{m_1\times n_2}$, $R\in\cU^{m_2\times n_1}$, for some positive integers $n_1,m_1,n_2,m_2$ with $m_1+m_2=m$,  $n_1+n_2=n$,    $n_1+m_1\ge 4$ and  $n_2+m_2\ge 4$:
	\begin{itemize}
		\item $1$-sum decomposition: $R=\bzero$ and $Q=\bzero$;
		\item $2$-sum decomposition: $R$ has rank $1$ and $Q=\bzero$; 
		\item $3$-sum decomposition -- case 1: $R$ and $Q$ have rank $1$;
		 \item $3$-sum decomposition -- case 2: $R$ has rank $2$, $Q=\bzero$. 
	\end{itemize}
	\end{itemize}
	Deciding if $A$ is a network matrix or its transpose, or (up to permutations of rows and columns)  $A_0$, and if not, finding a decomposition as above can be done in polynomial time. 
\end{theorem}

\begin{corollary}[Decomposition of unimodular hyeprgraphs]\label{c-decomp}
	Let $\cH\subseteq 2^V$ be a unimodular (nontrivial) irredundant Sperner hypergraph. Then $\cH$ is a network hypergraph, (isomorphic to) the hypergraph $\cH_0=\{\{1,4,5\},\{1,2,5\},\{2,3,5\},\{3,4,5\}\}$, has two identical vertices, has a hyperedge consisting of a singleton, has a vertex with degree $1$, or there exists a nontrivial partition $V_1\dot\cup V_2=V$ such that $\cH$ can be decomposed as follows:%\footnote{Note that the $1$-sum decomposition we may regarded as a special case of the $2$-decomposition, which can in turn be regarded as a  special case of the $3$-sum decomposition; but for clarity of presentation, we describe them separately.}:
	\begin{itemize}
		\item $1$-sum decomposition: 
		\begin{itemize}
			\item[(i)] $\cH_{V_1}\neq\emptyset$, $\cH_{V_2}\neq\emptyset$; 
		  %\item [(ii)] $|V_1|+|\cH_{V_1}|\ge4$, $|V_2|+|\cH_{V_2}|\geq 4$;
		  \item[(ii)] for all $H\in\cH$: either $H\subseteq V_1$ or $H\subseteq V_2$; 
		   \end{itemize}
        \item $2$-sum decomposition: there exists a set nonempty $S\subseteq V_1$ such that
        \begin{itemize}
        	\item[(i)]  $\cH_{V_1}\neq\emptyset$, $\cH(V_1,S)\neq\emptyset$, $\cH(V_1,S)^{V_2}\neq\{\emptyset\}$;
        	%\item[(ii)]  $|V_1|+|\cH_{V_1}| \ge 4$ and $|V_2|+|\cH_{V_2}\cup \cH(V_1,S)|\ge 4$;
        	\item[(ii)]  for all $H\in\cH$ with $H\cap V_1\neq\emptyset$ and $H\cap V_2\neq\emptyset$: $H\cap V_1=S$;
        \end{itemize}
		\item $3$-sum decomposition -- case 1: there exist two nonempty sets $S_1\subseteq V_1$ and $S_2\subseteq V_2$, such that 
		\begin{itemize}
		\item[(i)]  $\cH(V_1,S_1)\neq\emptyset$, $\cH(V_1,S_1)^{V_2}\neq\{\emptyset\}$, $\cH(V_2,S_2)\neq\emptyset$, $\cH(V_2,S_2)^{V_1}\neq\{\emptyset\}$;
		\item[(ii)]  $|V_1|+|\cH_{V_1}\cup \cH(V_2,S_2)|\ge 4$, $|V_2|+|\cH_{V_2}\cup \cH(V_1,S_1)|\ge 4$;
		\item[(iii)]  for all $H\in\cH$ with $H\cap V_1\not=\emptyset$ and $H\cap V_2\not=\emptyset$: either $H\cap V_1= S_1$ or $H\cap V_2=S_2$;
		 \end{itemize}
		\item $3$-sum decomposition -- case 2: there exist three nonempty disjoint sets $S_0,S_1,S_2\subseteq V_1$, such that 
		\begin{itemize}
		 \item[(i)]  $\cH_{V_1}\neq\emptyset$, $\cH(V_1,S_0\cup S_1)\neq\emptyset$, $ \cH(V_1,S_0\cup S_1)^{V_2}\neq\{\emptyset\}$, $\cH(V_1,S_0\cup S_2)\neq\emptyset$, $ \cH(V_1,S_0\cup S_2)^{V_2}\neq\{\emptyset\}$;
		 %\item[(ii)]  $|V_1|+|\cH_{V_1}|\ge4$, $|V_2|+|\cH_{V_2}\cup \cH(V_1,S_0\cup S_1)\cup \cH(V_1,S_0\cup S_2)|\ge4$;
		 \item[(ii)]  for all $H\in\cH$ with $H\cap V_1\neq\emptyset$ and $H\cap V_2\not=\emptyset$: either $H\cap V_1=S_0\cup S_1$, or $H\cap V_1= S_0\cup S_2$;
		  \end{itemize}
		\item $3$-sum decomposition -- case 3: there exist two nonempty disjoint sets $S_1,S_2\subseteq V_1$, such that 
		\begin{itemize}
			\item [(i)] $\cH_{V_1}\neq\emptyset$ and at least two of the following three  conditions hold: (1) $\cH(V_1,S_1)\neq\emptyset$ and $\cH(V_1,S_1)^{V_2}\neq\{\emptyset\}$, (2) $\cH(V_1,S_2)\neq\emptyset$, $\cH(V_1,S_2)^{V_2}\neq\{\emptyset\}$, (3)
			$\cH(V_1,S_1\cup S_2)\neq\emptyset$, $\cH(V_1,S_1\cup S_2)^{V_2}\neq\{\emptyset\}$;
			\item[(ii)]  $|V_1|+|\cH_{V_1}|\ge4$, $|V_2|+|\cH_{V_2}|\cup \cH(V_1,S_1)\cup\cH(V_1,S_2)\cup \cH(V_1,S_1\cup S_2)|\geq 4$; 
			\item [(iii)] for all $H\in\cH$ with $H\cap V_1\neq\emptyset$ and $H\cap V_2\not=\emptyset$: either $H\cap V_1=S_1$, $H\cap V_1=S_2$, or $H\cap V_1= S_1\cup S_2$.
		 \end{itemize}
	\end{itemize}
    Discovering if $\cH$ is a network hypergraph, or isomorphic to $\cH_0$, and if not finding a decomposition as above can be done in polynomial time. 
\end{corollary}
\proof
We consider only the $2$-sum and $3$-sum decomposition rules as  
the other cases are immediate from the corresponding cases in Theorem~\ref{t-decomp}. We define $V_1$ and $V_2$ to be the sets of columns of  the matrices $A^1$ and $A^2$ in Theorem~\ref{t-decomp}. Consider the $2$-sum case. Since $R$ has rank $1$, each nonzero row in $R$ is a copy of some row vector $e\in\{0,1\}^{V_1}$. Let $S=\{i\in V_1\mid e_i=1\}$. Then by the decomposition of the matrix $A$ in this case, all hyperedges $H\in\cH[A]$ such that $H\cap V_1\neq\emptyset$ and $H\cap V_2\neq\emptyset$, have $H\cap V_1=S$.

Consider next the $3$-sum--case 1 in Theorem~\ref{t-decomp}. Since $R$ and $Q$ have rank 1, each nonzero row in $R$ (resp., $Q$) is a copy of some row vector $e\in\{0,1\}^{V_1}$ (resp. $f\in\{0,1\}^{V_2}$). Let $S_1=\{i\in V_1\mid e_i=1\}$ and $S_2=\{i\in V_2\mid f_i=1\}$. Then by the decomposition of the matrix $A$ in this case, all hyperedges $H\in\cH[A]$ such that $H\cap V_1\neq\emptyset$ and $H\cap V_2\neq\emptyset$, have either $H\cap V_1=S_1$ or $H\cap V_2=S_2$.

Consider finally the $3$-sum--case 2  in Theorem~\ref{t-decomp}. Since $R$ has rank 2, there exist two rows $a,b\in\{0,1\}^{V_1}$ of $R$ that are linearly independent, such that any other row $c\in\{0,1\}^{V_1}$ of $R$ can be written as $c=\lambda_1a+\lambda_2b$, where $\lambda_1,\lambda_2\in\RR$.  We consider three subcases: 
\begin{itemize}
	\item [(I)] $a^Tb\ne 0$ and neither $a\ge b$ nor $b\ge a$: then  $(\lambda_1,\lambda_2)\in\{(0,0),(1,0),(0,1)\}$. Let $S_0=\{i\in V_1\mid a_i=1\text{ and } b_i=1\}$, $S_1=\{i\in V_1\mid a_i=1\text{ and } b_i=0\}$, and $S_2=\{i\in V_1\mid a_i=0\text{ and } b_i=1\}$. Then by the decomposition of the matrix $A$ in this case, all hyperedges $H\in\cH[A]$ such that $H\cap V_1\neq\emptyset$ and $H\cap V_2\neq\emptyset$, have either $H\cap V_1=S_0\cup S_1$, or $H\cap V_1=S_0\cup S_2$. This corresponds to case 2 of the $3$-sum decomposition in the corollary.
	\item [(II)] $a^Tb\ne 0$ and $a\ge b$ (resp., $b\ge a$): then $(\lambda_1,\lambda_2)\in\{(0,0),(1,0),(0,1),(1,-1)\}$ (resp., $(\lambda_1,\lambda_2)\in\{(0,0),(1,0),(0,1),(-1,1)\}$). Let $S_1=\{i\in V_1\mid a_i=1\text{ and } b_i=1\}$, and $S_2=\{i\in V_1\mid a_i=1\text{ and } b_i=0\}$ (resp., $S_2=\{i\in V_1\mid a_i=0\text{ and } b_i=1\}$). Then by the decomposition of the matrix $A$ in this case, all hyperedges $H\in\cH[A]$ such that $H\cap V_1\neq\emptyset$ and $H\cap V_2\neq\emptyset$, have either $H\cap V_1=S_1$, $H\cap V_1=S_2$, or $H\cap V_1=S_1\cup S_2$.  This corresponds to case 3 of the $3$-sum decomposition in the corollary.	
    \item [(III)] $a^Tb=0$: then $\lambda_1,\lambda_2\in\{0,1\}$. Let $S_1=\{i\in V_1\mid a_i=1\text{ and } b_i=0\}$, and $S_2=\{i\in V_1\mid a_i=0\text{ and } b_i=1\}$. Then by the decomposition of the matrix $A$ in this case, all hyperedges $H\in\cH[A]$ such that $H\cap V_1\neq\emptyset$ and $H\cap V_2\neq\emptyset$, have either $H\cap V_1=S_1$, $H\cap V_1=S_2$, or $H\cap V_1=S_1\cup S_2$. This corresponds to case 3 of the $3$-sum decomposition in the corollary.
\end{itemize}
Finally, we verify the boundary conditions. For the $1$-sum case, $\cH_{V_1},\cH_{V_2}$ are nonempty because of the requirement that $m_1,m_2\ge 1$ in Theorem~\ref{t-decomp}. Consider next the $2$-sum case. $\cH_{V_1}\neq\emptyset$ follows from the requirement that $m_1\ge 1$ in Theorem~\ref{t-decomp}. If $\cH(V_1,S)^{V_2}=\{\emptyset\}$ then, by the simplicity of $\cH$, we would have $\cH(V_1,S)=\{S\}\in\cH_{V_1}$, which would imply that we are also in the $1$-sum case. Consider next the $3$-sum case 1. $\cH(V_1,S_1)\ne\emptyset$ and $\cH(V_2,S_2)\ne\emptyset$ follow, respectively, from the requirements that $m_1\ge 1$ and $m_2\ge 1$ in Theorem~\ref{t-decomp}.  If either $\cH(V_1,S_1)^{V_2}=\{\emptyset\}$ or $\cH(V_2,S_2)^{V_1}=\{\emptyset\}$ then, again by the simplicity of $\cH$, we  are in the $1$-sum or the $2$-sum cases.
Next consider the $3$-sum--case 2. Note that $\cH_{V_1}\neq\emptyset$, $\cH(V_1,S_0\cup S_1)\neq\emptyset$ and $\cH(V_1,S_0\cup S_2)\neq\emptyset$ in this case. If either $\cH(V_1,S_0\cup S_1)^{V_2}=\{\emptyset\}$ (resp., $\cH(V_1,S_0\cup S_2)^{V_2}=\{\emptyset\}$) then by the simplicity of $\cH$, we would have $\cH(V_1,S_0\cup S_1)=\{S_0\cup S_1\}\in\cH_{V_1}$ (resp., $\cH(V_1,S_0\cup S_2)=\{S_0\cup S_2\}\in\cH_{V_1}$), implying that we are in either the $1$-sum or the $2-$sum cases. A similar argument applies for $3$-sum--case 3. 
\qed

\begin{figure}[htb!]
	\centering 
	\includegraphics[scale=0.5]{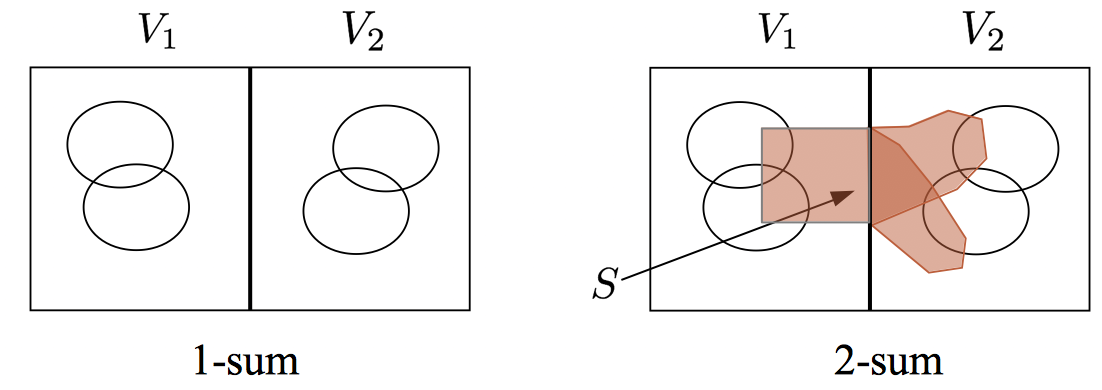} 
	\caption{Decomposing a unimodular hypergraph: $1$ and $2$-sums.}
	\label{f1}
\end{figure}

\begin{figure}[htb!]
	\centering 
	\includegraphics[scale=0.5]{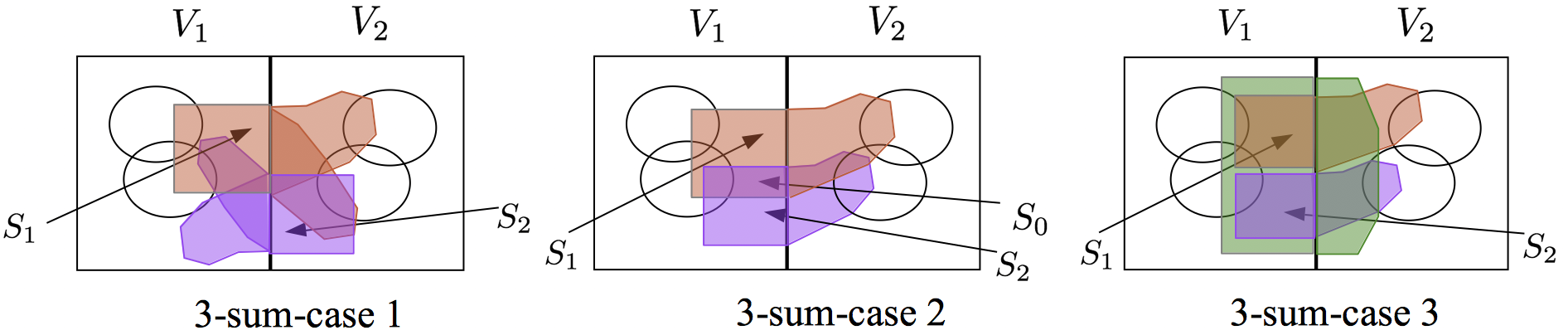} 
	\caption{Decomposing a unimodular hypergraph: $3$-sum.}
		\label{f2}
\end{figure}

A schematic illustration of these decomposition rules is given in Figures~\ref{f1} and~\ref{f2}.
%For completeness, the original decomposition in \cite{S80,T92} is given in Appendix~\ref{a:decomp}. 
\begin{remark}\label{r1}
We note that the boundary condition (ii) in the $3$-sum--case 1 is essential, since without insisting on this condition, any hypergraph can be decomposed according to the $3$-sum--case 1 rule (take any $v\in V$ and $H\in\cH$ such that $v\in H$, and let $V_1=S_1=\{v\}$, $V_2=V\setminus\{v\}$ and $S_2=H\setminus\{v\}$). Similarly, our analysis in the $3$-sum--case 3 uses condition (ii). However, a similar condition is not needed for all other cases. 
\end{remark}

\section{Decomposition of the hypergraph transversal problem}
In the following, we show how to decompose the hypergraph transversal problem for a unimodular hypergraph $\cH$, given the decomposition of $\cH$ as in Corollary~\ref{c-decomp}. Such a decomposition yields naturally a recursive algorithm: each non-leaf node of the recursion tree is responsible for computing the dual of a unimodular  hypergraph,  while leaves involve the computation of the dual of a network hypergraph or the hypergraph $\cH_0$. 
To ensure that the overall running time is polynomial, we need to bound the number of nodes of the recursion tree and the local computation time at each node, which consists of the time required for computing the decomposition and the time for combining the outputs from the recursive calls into the final output at the node. We will measure the ``volume" of each subproblem to compute $\Tr(\cH)$ by $\mu(\cH)=nmk=n(\cH)m(\cH)k(\cH)$. We let $T(\mu)$ be the number of nodes of the recursion subtree rooted at a node of volume $\mu$, and let $L_1(\mu)$ and $L_2(\mu)$ be respectively the local computation time for the decomposition and combining the outputs at a node of volume $\mu$. We
stop the recursion when either $m=m(\cH)$, $n=n(\cH)$ or $k=k(\cH)$ drops below some constant $C$, in which case the hypergraph transversal problem can be solved in $\poly(n,m)$ time using a simple procedure, such as {\it Berge Multiplication}~\cite[Chapter 2]{B89} for $n(\cH)\le C$ or $n(\cH)\le C$, and the methods in \cite{BI95,GK99} for $k(\cH)\le C$, which also show that the condition  $k(\cH)\le C$ can be checked in $\poly(n,m)$ time.   

We will show by induction (on $\mu\ge 1$) that $T(\mu) \le \mu$. We also show that $L_2(\mu)= O(\mu^{c})$ for some constant $c\ge 1$.  Since $L_1(\mu)=\poly(n,m)$ \cite[Chapter 20]{S86}, it would follow then that the total time to compute $\Tr(\cH)$ is at most $O(\mu^{1+c})+\poly(\mu)$, which is polynomial in $n$, $m$, and $k$. This would give a {\it total} polynomial-time algorithm for computing $\Tr(\cH)$ which can be converted into an {\it incremental} polynomial-time algorithm by standard methods \cite{KBEGM-SIDMA05,BEGM09}. Thus, we shall assume in the sequel that $n,m,k$ are larger than any desired constant $C$.

Without loss of generality we assume that the input hypergraph is Sperner and irredundant, and this assumption is maintained for all hypergraphs arising as inputs to the recursive subproblems. 
We may also assume that $\cH$ has neither a singleton hyperedge nor a vertex of degree 1 (i.e., contained in exactly one hyperedge). Indeed, if $\cH$ contains a singleton hyperedge $H=\{v\}$, then by the Sperner property, no other hyperedge of $\cH$ contains $v$.  In this case, and also in the case when $\cH$ has a vertex $v$ contained exactly in one hyperedge $H\in\cH$, $\Tr(\cH)$ can be computed as follows:
\begin{equation}
\label{e10-}
\Tr(\cH)=\Tr((\cH\setminus\{H\})\ \dot{\cup}\ \{H\})=\Tr((\cH\setminus\{H\}){\wedge} \Tr(\{H\}), 
\end{equation}
where $\Tr(\{H\}=\{\{w\} \mid w \in H\}$.  
By Proposition~\ref{prop1} (iii), $|\Tr((\cH\setminus\{H\})|\le k(\cH)$ and thus, $\mu(\cH')\le(n-1)(m-1)k\leq\mu(\cH)-1$. Thus, we get by induction that $T(\mu(\cH))\le 1+T(\mu(\cH'))\le \mu(\cH)$. 
Moreover, by \raf{e10-}, $\Tr(\cH)$ can be computed from $\Tr((\cH\setminus\{H\})$ in $L_2(\mu)=\poly(n,m,k)$ time.

Finally, we may also assume that $\cH$ does not have two identical vertices. Indeed, if it has two such vertices $v,v'$ then we can reduce the problem by calling the algorithm on the hypergraph $\cH'=\{H\setminus\{v'\}~|~H\in\cH\}$ instead of $\cH$. Then the dual of $\cH$ can be obtained as follows: 
\begin{equation}
\label{e11-}
\Tr(\cH)=\Tr(\cH')\dot{\cup}\{(T\setminus\{v\}\cup\{v'\})~|~T\in\Tr(\cH'),~v\in T\}.
\end{equation}
 Note that \raf{e11-} implies that $k(\cH')\le k(\cH)$ and hence
$\mu(\cH')\le (n-1)mk\le \mu(\cH)-1$.  Thus, in this case, we get the recurrence $T(\mu)\le 1+T(\mu-1)$, which again gives by induction on $\mu\ge 1$ that
%\begin{align}\label{t0}
$T(\mu)\le 1+(\mu-1)\le \mu.$
%\end{align}
Moreover,  by \raf{e11-}, $\Tr(\cH)$ can be computed from $\Tr(\cH')$ in  $\poly(n,m,k)$ time.   

\medskip

We will use the following simple facts in our analysis of the running time of the algorithm.

\begin{fact}\label{fact1}
	Let $\alpha,\beta,N, M$ be positive integers such that $\alpha\le N/2$ and $\beta\le M/2$. Consider the maximization problem:
	\begin{align*}\label{m1}
	z^*=\max& \quad x_1y_1+x_2y_2\\
	{\rm s.t.}&\quad  x_1+x_2 \le N,\\
	&\quad y_1+y_2 \le M,\\
	&\quad x_1,x_2\ge \alpha,\\
	&\quad y_1,y_2\ge \beta,\\
	&\quad x_1,x_2,y_1,y_2\in\ZZ.
	\end{align*} 
	Then $z^*=\alpha\beta+(N-\alpha)(M-\beta)$. 
\end{fact}	
\proof
Let $(x^*_1,x_2^*,y^*_1,y_2^*)$ be an optimal solution. Clearly, $x_2^*=N-x_1^*$ and $y_2^*=M-y_1^*$. Without loss of generality assume that $x_1^*\ge\frac{N}{2}$. If $y_1^*<M-\beta$, then $(x^*_1,N-x_1^*,y_1^*+1,M-y_1^*-1)$ is also an optimal solution since $$x^*_1(y^*_1+1)+(N-x^*_1)(M-(y^*_1+1))=x^*_1y^*_1+(N-x^*_1)(M-y^*_1)+2x^*_1-N\ge x^*_1y^*_1+(N-x^*_1)(M-y^*_1).$$
Thus we conclude in this case that $(x^*_1,N-x_1^*,M-\beta,\beta)$ is also an optimal solution. 
%On the other hand, if $x^*_1\le\frac{N}{2}$ and $y^*_1\ge\beta$ then $(x^*_1,N-x_1^*,y_1^*-1,M-y_2^*)$ is also an optimal solution  since
%$$x^*(y^*-1)+(N-x^*)(M-(y^*-1))=x^*y^*+(N-x^*)(M-y^*)-2x^*+N\ge x^*y^*+(N-x^*)(M-y^*).$$
%Thus we conclude in this case that $(x^*,\beta)$ is also an optimal solution. 
A symmetric argument shows that $(N-\alpha,\alpha,M-\beta,\beta)$ is an optimal solution of the maximization problem.
\qed

\begin{fact}\label{fact2}
	Let $x_i,y_i$, for $i=1,\ldots,h$, and $M$ be positive integers such that $\sum_{i=1}^hx_iy_i\le M$. Then 
	%\begin{equation}\label{ep-}
	$\sum_{i=1}^h(x_i+y_i)\le M+h.$
	%\end{equation}
\end{fact}	
\proof
For $i=1,\ldots,h$, let $\alpha_i=x_iy_i$. Note that the function $f(x_i)=x_i+\frac{\alpha}{x_i}$ is convex in $x_i>0$, and hence $\max \{x_i+y_i\mid x_iy_i=\alpha_i,~x_i\ge1,y_i\ge1\}$ is achieved at the boundary $(x_i,y_i)=(1,\alpha_i)$ or  $(x_i,y_i)=(\alpha_i,1)$. The claim follows by summing the inequality $x_i+y_i\le \alpha_i+1$ over $i=1,\ldots,h$.
\qed

 \subsection{$1$-sum decomposition}
 Given a nontrivial partition $V_1\dot\cup V_2=V$ such that  for all $H\in\cH$: either $H\subseteq V_1$ or $H\subseteq V_2$, we have the following decomposition of the dual hypergraph by  Proposition~\ref{prop2}:
 \begin{equation}\label{e1}
    \Tr(\cH)=\Tr(\cH_{V_1}\dot\cup\cH_{V_2})=\Tr(\cH_{V_1})\dot\wedge\Tr(\cH_{V_2}).
 \end{equation}
 Note that both $\cH_{V_1}$ and $\cH_{V_2}$ are unimodular. Thus in this case we get the recurrence:
  \begin{align}\label{rec1}
 T(\mu)\le 1+T(\mu_1)+T(\mu_2),
 \end{align}
 where $\mu=\mu(\cH)$, $\mu_1=\mu(\cH_{V_1})$, and $\mu_2=\mu(\cH_{V_2})$.
 Let $n_1=|V_1|$, $m_1=|\cH_{V_1}|$, $k_1=|\Tr(\cH_{V_1})|$,  $n_2=|V_2|$, $m_2=|\cH_{V_2}|$, and $k_2=|\Tr(\cH_{V_2})|$. Note that $n_1,n_2,m_1,m_2\ge1$, $n_1+n_2=n$ and $m_1+m_2=m$ by the assumptions of the $1$-sum case (Corollary~\ref{c-decomp}) and hence $\mu_1,\mu_2\ge 1$. By Proposition~\ref{prop1}(iii), $k_1\le k$ and $k_2\le k$ (in fact, $k_1k_2=k$). Thus, it follows by Fact~\ref{fact1} and $n\ge 2$, $m\ge 2$ and $k\ge 1$ that
 \begin{align}\label{v1}
 \mu_1+\mu_2&=n_1m_1k_1 +n_2m_2k_2\le (n_1m_1+n_2m_2)k\nonumber\\
 &\le [1+(n-1)(m-1)]k=[2+nm-n-m]k\le (nm-2)k\le nmk-2=\mu-2.
 \end{align}
%where the first inequality follows by Proposition~\ref{p3-}. 
  It follows by induction from \raf{rec1} that 
  \begin{align}\label{t1}
    T(\mu)\le 1+\mu_1+\mu_2\le 1+(\mu-2)\leq\mu.
  \end{align}
 Note also that $\Tr(\cH)$ can be computed from $\Tr(\cH_{V_1})$ and $\Tr(\cH_{V_2})$ using~\raf{e1} in time $L_2(\mu)=\poly(n,m,k)$.
 %=O(k_1k_2n)=O(\mu)$ time.
\subsection{$2$-sum decomposition}
Given a nontrivial partition $V_1\dot\cup V_2=V$ and a nonempty set $S\subseteq V_1$ such that for all $H\in\cH$ with $H\cap V_1\neq\emptyset$ and $H\cap V_2\neq\emptyset$: $H\cap V_1=S$, we have the following decomposition of the dual hypergraph by  Proposition~\ref{prop2}:
\begin{equation}\label{e2}
\Tr(\cH)=\Tr(\cH_{V_1}\dot\cup\cH_{V_2\cup S})=\Tr(\cH_{V_1})\wedge\Tr(\cH_{V_2\cup S}),
\end{equation}
 as $\cH=\cH_{V_1}\dot{\cup}\cH_{V_2\cup S}$
(note that $\cH_{V_1}\cap \cH_{V_2\cup S}=\emptyset$ since $\cH$ is Sperner and  that both $\cH_{V_1}$ and $\cH_{V_2\cup S}$ are unimodular).
Thus in this case we get the recurrence:
\begin{align}\label{rec1-}
T(\mu)\le 1+T(\mu_1)+T(\mu_2),
\end{align}
where $\mu=\mu(\cH)$, $\mu_1=\mu(\cH_{V_1})$, and $\mu_2=\mu(\cH_{V_2\cup S})$.
Let $n_1=n(\cH_{V_1})=|V_1|$, $m_1=|\cH_{V_1}|$, $k_1=|\Tr(\cH_{V_1})|$,  $n_2=n(\cH_{V_2\cup S})=|V_2|+|S|$, $m_2=|\cH_{V_2\cup S}|$, and $k_2=|\Tr(\cH_{V_2\cup S})|$. Note that $n_1,n_2,m_1,m_2\ge1$ by the assumptions of the $2$-sum case (Corollary~\ref{c-decomp}) and hence $\mu_1,\mu_2\ge 1$. Then
\begin{align}\label{v2}
\mu_1+\mu_2&=n_1m_1k_1 +n_2m_2k_2\le (n-1)(m_1+m_2)k=(n-1)mk\le \mu-1,
\end{align}
where $k_1\le k$ and $k_2\le k$ by Proposition~\ref{prop1} (iii). It follows by induction from \raf{rec1-} that 
\begin{align}\label{t2}
T(\mu)\le 1+\mu_1+\mu_2\le 1+(\mu-1)=\mu.
\end{align}
Note that $\Tr(\cH)$ can be computed from $\Tr(\cH_{V_1})$ and $\Tr(\cH_{V_2\cup S})$ using~\raf{e2} in time $L_2(\mu)=\poly(n,m,k)$.
%O(k_1k_2mn^2)=O(k^2mn^2)=O(\mu^2)$ time\footnote{For simplicity, we do not attempt to optimize the bounds on the running time}.
\subsection{$3$-sum decomposition - case 1}
Assume we are given a nontrivial partition $V_1\dot\cup V_2=V$ and two nonempty sets $S_1\subseteq V_1$ and $S_2\subseteq V_2$, such that  for all $H\in\cH$ with $H\cap V_1\not=\emptyset$ and $H\cap V_2\not=\emptyset$: either $H\cap V_1= S_1$ or $H\cap V_2=S_2$. Let $n_1=|V_1|$, $n_2=|V_2|$, $m_1=|\cH_{V_1}\cup \cH(V_2,S_2)|$ and $m_2=|\cH_{V_2}\cup \cH(V_1,S_1)|$. It is also assumed in this case that $n_1,n_2,m_1,m_2\ge 1$, $n_1+n_2=n$, $m_1+m_2=m$, $n_1+m_1\ge 4$ and $n_2+m_2\ge 4$, and that $\cH(V_1,S_1)$ and $\cH(V_2,S_2)$ are not empty.%; otherwise, we are in the $1$-sum or $2$-sum cases. 

\medskip

We consider two cases:

\medskip

\noindent {\bf Case I}:  there is no hyperedge $H\in\cH$ such that $H\subseteq S_1\cup S_2$. Note that this, together with assumption (i) of the $3$-sum--case 1 in Corollary~\ref{c-decomp}, implies that $S_1\subset V_1$ and $S_2\subset V_2$. In this case, we have the following decomposition of the dual hypergraph:
	\begin{equation}\label{e3-1}
	\Tr(\cH)=\Tr(\cH_{V_1\cup S_2}\dot{\cup}\cH_{V_2\cup S_1})=\Tr(\cH_{V_1\cup S_2})\wedge\Tr(\cH_{V_2\cup S_1}),
	\end{equation}
	(Note by assumption that $\cH_{V_1\cup S_2}\cap \cH_{V_2\cup S_1}=\emptyset$.)
	Thus in this case we get the recurrence:
	\begin{align}\label{rec1--}
	T(\mu)\le 1+T(\mu_1)+T(\mu_2),
	\end{align}
	where $\mu=\mu(\cH)$, $\mu_1=\mu(\cH_{V_1\cup S_2})$, and $\mu_2=\mu(\cH_{V_2\cup S_1})$.
	Let $n_1'=n(\cH_{V_1\cup S_2})=|V_1|+|S_2|$, $m_1=|\cH_{V_1\cup S_2}|$, $k_1=|\Tr(\cH_{V_1\cup S_2})|$,  $n_2'=n(\cH_{V_2\cup S_1})=|V_2|+|S_1|$, $m_2=|\cH_{V_2\cup S_1}|$, and $k_2=|\Tr(\cH_{V_2\cup S_1})|$. %Note that $n_1',n_2',m_1,m_2\ge1$ by the assumptions of the $3$-sum case (Corollary~\ref{c-decomp}) and hence $\mu_1,\mu_2\ge 1$. 
	Then
	\begin{align}\label{v3}
	\mu_1+\mu_2&=n_1'm_1k_1 +n_2'm_2k_2\le (n-1)(m_1+m_2)k\le\mu-1,
	\end{align}
	where $k_1\le k$ and $k_2\le k$ by Proposition~\ref{prop1} (iii). It follows by induction from \raf{rec1--} that 
	\begin{align}\label{t3}
	T(\mu)\le 1+\mu_1+\mu_2\le \mu.
	\end{align}
	Note that $\Tr(\cH)$ can be computed from $\Tr(\cH_{V_1\cup S_2})$ and $\Tr(\cH_{V_2\cup S_1})$ using ~\raf{e3-1} in time $L_2(\mu)=\poly(n,m,k)$.
	%O(k_1k_2mn^2)=O(k^2mn^2)=O(\mu^2)$ time.
	\medskip
	
\noindent {\bf Case II}:  there is a hyperedge $H_0\in\cH$ such that $H_0\subseteq S_1\cup S_2$. Note that $H_0\cap S_1\neq\emptyset$ and $H_0\cap S_2\neq\emptyset$ since otherwise by the simplicity of $\cH$ we are in the $2$-sum case. Without loss of generality, assume that $H_0\cap V_1=S_1$ and $H_0\cap V_2\subseteq S_2$. We assume that $\cH(V_1,S_1)$ and $\cH(V_2,S_2)$ are not empty; otherwise, we are in the $1$-sum or $2$-sum cases. Given these assumptions, we use the following decomposition of the dual hypergraph:
\begin{equation}\label{e3-2}
\Tr(\cH)=\Tr(\cH_1{\cup}\cH_2)=\Tr(\cH_1)\wedge\Tr(\cH_2),
\end{equation}
where $\cH_1=\cH_{V_1}\cup\cH(V_2,S_2)\cup\{H_0\}$ and $\cH_2=\cH_{V_2}\dot \cup\cH(V_1,S_1)$.
Note that $H_0$ is the the only hyperedge that belongs to both $\cH_1$ and $\cH_2$. Note also that neither $\cH_1$ nor $\cH_{2}$ may be a projection of $\cH$ (i.e., of the form  $\cH_S$ for some $S\subseteq V$) since there are hyperedges $H\subseteq S_1\cup S_2$ that may not be included in $\cH_1$ and $\cH_2$. Hence, Proposition~\ref{prop1} {\it cannot} be used to bound the sizes of $\Tr(\cH_1)$ and $\Tr(\cH_2)$. Nevertheless, due to the special structure of the decomposition in this case, we can use the bounds given in Lemma~\ref{cl1} below instead.  Let $\bar \cH_1\subseteq 2^{V_1\cup \{v_2\}}$ (resp., $\bar \cH_2\subseteq 2^{V_2\cup \{v_1\}}$) be the hypergraph obtained from $\cH_1$ (resp.,  $\cH_2$) by replacing $S_2$ (resp., $S_1$) by a {\it new} single vertex $v_2$ (resp., $v_1$), that is, 
\begin{align*}
\bar\cH_1=\cH_{V_1}\cup \bar \cH(V_2,S_2)\cup\{\bar H_0\},\qquad
\bar\cH_2=\cH_{V_2}\cup\bar \cH(V_1,S_1), 
\end{align*}
where $\bar \cH(V_2,S_2)=\{(H\setminus S_2)\cup\{v_2\}\mid H\in\cH(V_2,S_2)\}$, $\bar H_0=(H_0\setminus S_2)\cup\{v_2\}$,  and $\bar \cH(V_1,S_1)=\{(H\setminus S_1)\cup\{v_1\}\mid H\in\cH(V_1,S_1)\}$.

\begin{lemma}\label{cl0}
If $\cH$ is unimodular, then both $\bar \cH_1$ and $\bar \cH_2$ are unimodular.
\end{lemma}
\proof
Let $v$ be an arbitrary vertex in $H_0\cap S_2$. Then the (hyperedge-vertex) incidence matrix of the hypergraph $\bar\cH_1$ is a submatrix of that of $\cH$, with rows restricted to $\cH_{V_1}\cup\cH(V_2,S_2)\cup\{H_0\}$, and columns restricted to $V_1\cup\{v\}$. This shows that this submatrix is totally unimodular. A similar argument shows that $\bar \cH_2$ is also unimodular.
\qed
		
\begin{lemma} \label{cl1}
	$|\Tr(\bar\cH_1)|\le|\Tr(\cH)|$ and $|\Tr(\bar\cH_2)|\le|\Tr(\cH)|$.
\end{lemma}	   
\proof
We prove the claim that $|\Tr(\bar\cH_1)|\le|\Tr(\cH)|$;  the other claim can be proved similarly. 

It is enough to show that for  every minimal transversal $T\in \Tr(\bar\cH_1)$, there is a minimal transversal $T'\in\Tr(\cH)$ such that for any distinct $T_1,T_2\in\Tr(\bar\cH_1)$, $T_1'$ and $T_2'$ are distinct. 

Let $\cT_1=\{T\in\Tr(\bar \cH_1): v_2\not\in T\}$ and $\cT_2=\Tr(\bar \cH_1)\setminus\cT_1$. Consider first $T\in\cT_1$. By assumption  $T\cap S_1\neq\emptyset$ since $T$ has a nonempty intersection with $\bar H_0$. It follows that the only hyperedges of $\cH$ having empty intersection with $T$ are those in $\cH_{V_2}$. Note that none of these hyperedges are contained in $S_2$ since $\cH$ is Sperner. This implies that  $\cH_{V_2}^{V_2\setminus S_2}\neq\{\emptyset\}$ and therefore $\Tr(\cH_{V_2}^{V_2\setminus S_2})\neq\emptyset$. Let $T''$ be an arbitrary minimal transversal in $\Tr(\cH_{V_2}^{V_2\setminus S_2})$. Then it is easy to see that $T'=T\cup T''$ is in $\Tr(\cH)$.

Consider now $T\in\cT_2$. By the minimality of $T$, there is a hyperedge $H\in\bar\cH(V_2,S_2)\cup\{\bar H_0\}$ such that $H\cap T=\{v_2\}$. Furthermore, for every $v\in T\setminus\{v_0\}$, there is an $H\in\cH_{V_1}$ such that $T\cap H=\{v\}$. Let $\cH(T)=\{H\in\cH\mid H\cap T\setminus\{v_2\}=\emptyset\}$ and note that $\cH(T)^{V_2}$ is nontrivial. Pick $T''\in\Tr(\cH(T)^{V_2})$ arbitrarily. 
Then it is easy to see that $T'=T\cup T''$ is in $\Tr(\cH)$. 

Finally, note that for any distinct $T_1,T_2\in\cT_1$ (resp., $T_1,T_2\in\cT_2$), the constructed minimal transversals $T_1',T_2'\in\T(\cH)$ are distinct.  Moreover, for $T_1\in\cT_1$ and $T_2\in\cT_2$, $T_1'$ and $T'_2$ are distinct because $T'_1\cap S_2=\emptyset$ while $T_2'\cap S_2\not=\emptyset$.
\qed

To compute~\raf{e3-2}, we find $\Tr(\bar\cH_1)$ and $\Tr(\bar \cH_2)$, recursively. Then  $\Tr(\cH_1)$ and $\Tr(\cH_2)$ are given by the following claim.
\begin{lemma}\label{cl2}
	Let $\cT_1=\{T\in\Tr(\bar \cH_1)\mid  v_2\not\in T\}$, $\cT_2=\{T\in\Tr(\bar \cH_1)\mid  v_2\in T,~S_1\cap T=\emptyset\}$, $\cT_3=\Tr(\bar \cH_1)\setminus(\cT_1\cup\cT_2)$,
	$\cT_1'=\{T\in\Tr(\bar \cH_2)\mid  v_1\not\in T\}$ and $\cT_2'=\Tr(\bar \cH_2)\setminus\cT_1'$. Then
	\begin{align}\label{e3-3}
	\Tr(\cH_1)&=\cT_1\dot{\cup}\{(T\setminus \{v_2\})\cup\{v\}\mid v\in H_0\cap S_2\text{ and }T\in\cT_2\}\dot\cup \{(T\setminus \{v_2\})\cup \{v\}\mid v\in S_2\text{ and }T\in\cT_3\},\\
	\Tr(\cH_2)&=\cT_1'\dot\cup \{(T\setminus \{v_1\})\cup\{v\}\mid v\in S_1\text{ and }T\in\cT_2'\}.\label{e3-4}
	\end{align}
\end{lemma}
\proof
Let us prove \raf{e3-3}, since the proof  of \raf{e3-4} is similar. Suppose $T\in\Tr(\cH_1)$. If $T\cap S_2=\emptyset$ then (it is easy to see that) $T\in\cT_1$.
If $T\cap S_2\neq\emptyset$ then by minimality of $T$, $|T\cap S_2|=1$; let $T\cap S_2=\{v\}$. If $T\cap S_1=\emptyset$ then necessarily $v\in H_0$, in which case $(T\setminus\{v\})\cup\{v_2\}\in\cT_2$; otherwise $v$ can be any element in $S_2$, and hence, $(T\setminus\{v\})\cup\{v_2\}\in\cT_3$. On the other direction, if $T\in\cT_1$ then clearly $T\in\Tr(\cH_1)$; if $T\in\cT_2$ then 
$T\cap\bar H_0=\{v_2\}$ which implies that $(T\setminus \{v_2\})\cup\{v\}\in \Tr(\cH_1)$ for every $v\in H_0\cap S_2$; finally, if $T\in\cT_3$ then 
there is a hyperedge $H\in\bar\cH(V_2,S_2)$ such that $H\cap T=\{v_2\}$, which implies in turn that $(T\setminus \{v_2\})\cup\{v\}\in \Tr(\cH_1)$.
\qed

Note that $\Tr(\cH)$ can be computed from $\Tr(\bar\cH_1)$ and $\Tr(\bar\cH_2)$ using ~\raf{e3-2} and Lemma~\ref{cl2} in time $L_2(\mu)=\poly(n,m,k)$.

Let $n_1'=n(\bar{\cH_1})=n_1+1$, $m_1'=|\bar\cH_{1}|\in\{m_1, m_1+1\}$, $k_1=|\Tr(\bar\cH_{1})|$,  $n_2'=n(\bar{\cH_2})=n_2+1$, $m_2=|\bar\cH_{2}|$, and $k_2=|\Tr(\bar\cH_{2})|$. By the decomposition, $n_1'+n_2'=n+2$ and $m_1'+m_2= m+1$, and by Lemma~\ref{cl1}, $k_1\le k$ and $k_2\le k$. Note that $n_1',n_2'\ge 2$, $m_1',m_2\ge1$, $n_1+m_1 \ge 4$, and $n_2+m_2\ge 4$, by the assumptions of the $3$-sum case 1 in Corollary~\ref{c-decomp}. 

\medskip

We consider 3 subcases.

\medskip

\noindent {\bf Case II-I}: $2\le n_1'\le 3$. Then a simple procedure will be used to compute $\Tr(\bar\cH_1)$, and hence we need only to recurse on $\bar\cH_2$, giving the simpler recurrence: $T(\mu)\le 2+T(\mu_2)$. Note that $m_2\le m-2$ since $n_1\le 2$ implies $m_1\ge 2$ and hence $m_2=m-m_1\le m-2$. Since $\mu_2=n_2'm_2k_2\le n(m-2)k\le\mu-2$, we get by induction that 
\begin{align}\label{t4}
T(\mu)\le 2+\mu_2\le \mu.
\end{align}

\medskip

\noindent {\bf Case II-II}: $n_2'=2$. Then a simple procedure will be used to compute $\Tr(\bar\cH_2)$, and hence we need only to recurse on $\bar\cH_1$, giving the simpler recurrence: $T(\mu)\le 2+T(\mu_1)$. As above, $m_1\le m-3$ implying that $\mu_1=n_1'm_1'k_2\le n(m-2)k\le\mu-2$, and giving by induction again that $T(\mu)\le \mu.$

\medskip

\noindent {\bf Case II-III}: $n_1'\ge 4$ and $n_2'\ge 3$. We first note that $m_1',m_2\ge 2$. Indeed, if $m_1'=1$ (resp., $m_2=1$), then $\cH_{V_1}=\emptyset$ and $\cH(V_2,S_2)=\{H_0\}$ (resp., $\cH_{V_2}=\emptyset$ and $\cH(V_1,S_1)=\{H_0\}$). Since we assume that $\cH$ does not have identical vertices, we must have $n_1=1$ (resp., $n_2=1$). In either case we get a contradiction to the boundary assumtpions (ii) of the $3$-sum--case 1 in Corollary~\ref{c-decomp}.
Lemmas~\ref{cl0} and~\ref{cl2} imply that, in this case, we get the recurrence:
\begin{align}\label{rec1---}
T(\mu)\le 1+T(\mu_1)+T(\mu_2),
\end{align}
where $\mu=\mu(\cH)$, $\mu_1=\mu(\bar\cH_{1})$, and $\mu_2=\mu(\bar\cH_{2})$.

Then by Fact~\ref{fact1}, applied with $x_1=n_1'$, $y_1=m_1'$,  $x_2=n_2'$, $y_2=m_2$, $N=n+2$, $M=m+1$, $\alpha=3$ and $\beta=2$, we get (as $n\ge 5$ and $m\ge 3$)
\begin{align}\label{v5}
\mu_1+\mu_2&=n_1'm_1'k_1 +n_2'm_2k_2\le(n_1'm_1'+n_2'm_2)k\le((n-1)(m-1)+6)k
\nonumber\\ 
&= nmk-(n+m-7)k \le\mu-1.
\end{align}
It follows by induction from \raf{rec1---} that 
\begin{align}\label{t5}
T(\mu)\le 1+\mu_1+\mu_2\le\mu.
\end{align}
%Note that $\Tr(\cH)$ can be computed from~\raf{e3-2},\raf{e3-3} and \raf{e3-4} in time $L(\mu)=O(k_1k_2n^3m)=O(k^2n^3m)=O(\mu^3)$ time.

\subsection{$3$-sum decomposition - case 2}
\label{section-3-sum-case2}

Let 
$\cH_1=\cH_{V_1}$ and $\cH_2=\cH_{V_2}$. 
By Corollary \ref{c-decomp}, we have three nonempty disjoint sets $S_0,S_1,S_2$ in $V_2$, and 
the following two families are nonempty:
\begin{eqnarray}
\cF_1&=&\{H \in \cH \mid H \cap V_1=S_0 \cup S_2, H \cap V_2\not=\emptyset\},\\
\cF_2&=&\{H \in \cH \mid H \cap V_1=S_0 \cup S_1, H \cap V_2\not=\emptyset\}. 
\end{eqnarray} 
Note that $V_1,V_2\not=\emptyset$, $\cH_1\not=\emptyset$, and 
$\cH$ can be partitioned in the following way.
\begin{equation}
\label{eq--aed1}
\cH=\cH_1 \dot{\cup} \cH_2 \dot{\cup} \cF_1 \dot{\cup} \cF_2, 
\end{equation}
where $\dot{\cup}$ denotes the disjoint union. 
For $i=0,1,2$, let 
\begin{equation}
\cT_i=\{T \in \Tr(\cH_1) \mid T\cap S_i \not=\emptyset, T \cap S_j=\emptyset\,\, (j\not=i)\},
\end{equation} 
and let 
\begin{equation}
\cT=\Tr(\cH_1) \setminus (\cT_0 \cup \cT_1 \cup \cT_2).
\end{equation}
%By the definition, $\cT_0$ consists of minimal transversals of $\cH_1$ that intersect at least two of $S_i$, and  
By definition, we have 
\begin{equation}
\Tr(\cH_1)=\cT \dot{\cup} \cT_0 \dot{\cup} \cT_1 \dot{\cup} \cT_2.  
\end{equation}
Let 
\begin{equation}
\cP=\cH_{S_0 \cup S_1\cup S_2}\,\,(=\{H \in \cH \mid H \subseteq S_0\cup S_1 \cup S_2\}). 
\end{equation}

We separately consider the following 4 cases.

\medskip
\noindent
{\bf Case I}: $\cP=\emptyset$. 

\smallskip

\noindent
{\bf Case II}: $\cP=\{S_0\cup S_1 \cup S_2\}$. 

\smallskip

\noindent
{\bf Case III}: $\cP\not= \emptyset, \{S_0\cup S_1 \cup S_2\}$ and  $\cT\not=\emptyset$. 

\smallskip

\noindent
{\bf Case IV}: $\cP\not= \emptyset, \{S_0\cup S_1 \cup S_2\}$ and  $\cT=\emptyset$. 
%\medskip
\noindent

\subsubsection*{Case I}
$\cH$ can be partitioned into $\cH_{1}$ and $\cH_{S_0\cup S_1\cup S_2\cup V_2}$, i.e., 
$\cH=\cH_{1} \dot{\cup}\cH_{S_0\cup S_1\cup S_2\cup V_2}$ and $\cH_{1}, \cH_{S_0\cup S_1\cup S_2\cup V_2}\not=\emptyset$. 
Since $\Tr(\cH)=\Tr(\cH_{1}) \wedge \Tr(\cH_{S_0\cup S_1\cup S_2\cup V_2})$, we obtain $\Tr(\cH)$ by computing $\Tr(\cH_{1})$ and $\Tr(\cH_{S_0\cup S_1\cup S_2\cup V_2})$. 
Let $n_1=|V_1|$, $m_1=|\cH_{1}|$, $k_1=|\Tr(\cH_{1})|$, 
$n'_2=|S_0\cup S_1\cup S_2\cup V_2|$, 
$m'_2=|\cH_{S_0\cup S_1\cup S_2\cup V_2}|$, $k_2=|\Tr(\cH_{S_0\cup S_1\cup S_2\cup V_2})|$. 
Similar to the $2$-sum decomposition case, we can show that $T(\mu)\le \mu$ and the computation of $\Tr(\cH)$ can be done in time $L_2(\mu)=\poly(n,m,k)$. 

\subsubsection*{Case II}
We consider two cases: {\bf II-I}: $|\cH_{1}| \geq 2$ and {\bf II-II}: $|\cH_{1}|=1$. 

\medskip
\noindent
{\bf Case II-I}: $|\cH_{1}| \geq 2$. 
Let $\cG$ be a hypergraph obtained from $\cH_{S_0\cup S_1\cup S_2\cup V_2}$ by replacing $S_0, S_1$, and $S_2$ by new vertices $v_0$, $v_1$ and $v_2$, respectively. 
For any hyperedge $H \in \cH_{S_0\cup S_1\cup S_2\cup V_2}$, 
$H \cap S_i\not=\emptyset$ implies that $S_i \subseteq H$.
Thus $\cG$ is well-defined. 
Note that  $\Tr(\cH_{S_0\cup S_1\cup S_2\cup V_2})$ can be obtained from $\Tr(\cG)$ in polynomial time by replacing $v_i$ with any element in $S_i$. 
Since $\cH=\cH_{1} \cup\cH_{S_0\cup S_1\cup S_2\cup V_2}$,  we have $\Tr(\cH)=\Tr(\cH_{1}) \wedge \Tr(\cH_{S_0\cup S_1\cup S_2\cup V_2})$. We thus decompose $\cH$ into $\cH_1$ and $\cG$. Namely we compute $\Tr(\cH)$ from $\Tr(\cH_1)$ and $\Tr(\cG)$.  
Since $|\Tr(\cH_1)|, |\Tr(\cG)|\leq |\Tr(\cH)|\,(=k)$, 
this can be done in time $L_2(\mu)=\poly(n,m,k)$. 

Let us next show that $T(\mu)\leq \mu$.
Let $n_1=|V_1|$, $m_1=|\cH_{1}|$, $k_1=|\Tr(\cH_{1})|$, 
$n'_2=|V_2|$+3, 
$m'_2=|\cG|$, and $k_2=|\Tr(\cG)|$.
%Note that $ S_0\cup S_1 \cup S_2 \subsetneq V_1$ holds, 
%since $S_0 \cup S_1 \cup S_2 \in \cH$, $|\cH_1|\geq 2$, 
%and $\cH$ is Sperner.  
Note that $\cH=\cH_{1} \cup\cH_{S_0\cup S_1\cup S_2\cup V_2}$, $\cH_{1}\cap \cH_{S_0\cup S_1\cup S_2\cup V_2}=\{S_0\cup S_1\cup S_2\}$. 
This, together with definition and the discussion above, implies  
that 
\begin{equation*}
1\leq n_1,n_2' \leq n, \ n_1 +n'_2=n+3, \ 
2\leq m_1, m'_2 \leq m-1, \ m_1 +m'_2=m+1, \ 
k_1,k_2 \leq k. 
\end{equation*}
Thus we have 
\begin{equation}
\label{eq3d8ge1-eq}
n_1m_1k_1 + n_2'm_2'k_2 \leq (n_1m_1+ n_2'm_2')k\leq (n(m-1)+6)k \leq nmk-1
\end{equation}
where Fact \ref{fact1} is used for the second inequality, 
and the third ineuqality is obtained by assuming that $n$ is at least $7$. 
It follows from \raf{eq3d8ge1-eq} that $T(\mu) \leq \mu$. 
We recall that $\Tr(\cH)$ is directly computed from $\cH$ 
if at least one of $n$, $m$, 
and $k$ is bounded by some constant $C$. 
Thus in case $n< 7$, we have $T(\mu)=1$, which also satisfies $T(\mu) \leq \mu$.

\medskip
\noindent
{\bf Case II-II}: $|\cH_{1}| =1$. In this case, we have $H_1=\{S_0 \cup S_1\cup S_2\}$. 
Therefore, the following lemma is satisfied. 
\begin{lemma}
	\label{lemma-sum3-2-2-1}
	Let $\cH$ be a hypergraph that satisfies \raf{eq--aed1}
	and $\cH_1=\{S_0 \cup S_1\cup S_2\}$. 
	Then we have $\Tr(\cH)=\{\{v\} \mid v\in S_0\} \dot\wedge \Tr(\cH_2) \ \dot\cup \ \Tr(H^{V\setminus S_0})$. 
\end{lemma} 

\noindent
\proof
From the definition, it is not difficult to see that $\Tr(\cH)\supseteq \{\{v\} \mid v\in S_0\} \dot\wedge \Tr(\cH_2) \ \dot\cup \ \Tr(H^{V\setminus S_0})$. 

For the converse inclusion, let $T\in \Tr(\cH)$. 
If $T \cap S_0=\emptyset$, then $T$ is contained in $\Tr(H^{V\setminus S_0})$. 
Assume next that $T \cap S_0 \not=\emptyset$. 
For any $i=0,1,2$ and  any hyperedge $H \in \cH$,  
$H \cap S_i\not=\emptyset$ inplies that $S_0 \subseteq H$. 
This means that $|T\cap S_0|=1$ and $T \cap S_i=\emptyset$ for $i=1,2$. Moreover, we have $T\cap V_2 \in \Tr(\cH_2)$, which completes the converse inclusion.  
\qed

Note that $\cH_2,\cH^{V\setminus S_0}\not=\{\emptyset\}$, and hence 
$\Tr(\cH_2),\Tr(H^{V\setminus S_0})\not=\emptyset$. 
Based on Lemma \ref{lemma-sum3-2-2-1}, we decompose $\cH$ into $\cH_2$ and $\cH^{V\setminus S_0}$. Namely, we compute $\Tr(\cH)$ from 
$\Tr(\cH_2)$ and $\Tr(H^{V\setminus S_0})$ in time $L_2(\mu)=\poly(n,m,k)$.  

Let $n'_1=|V_2|$, $m'_1=|\cH_2|$, $k'_1=|\Tr(\cH_2)|$, 
$n'_2=n-|S_0|$, $m'_2=|\cH^{V\setminus S_0}|$, and $k'_2=|\Tr(\cH^{V\setminus S_0})|$. 
Then we have $n'_1, n'_2\leq n-1$, $m'_1, m'_2\leq m$, and 
$k'_1, k_2'\leq k-1$ and $k_1'+k_2'\leq k$.
Thus we have
$n_1'm_1'k_1'+n_2'm_2'k_2'\leq (n-1)(m-1)k\leq nmk-1$, 
where the last inequality is obtained from $n\geq 3$. 
This implies that $T(\mu)\leq \mu$.

\subsubsection*{Case III}
For a set $W \subseteq V$, define $\eta(W)\subseteq \{0,1,2\}$ by
\[
\eta(W)=\{i\mid W \cap S_i\not=\emptyset\}.
\]
For $i=1,2$, let 
\begin{eqnarray}
\cF_i'&=&\Min(\cF_i^{V_2} \cup \cH_2).
\end{eqnarray} 
\begin{lemma}
	\label{lemma-maxk12}
	Let $\cH$ be a hypergraph that satisfies \raf{eq--aed1} and $\cP\not=\emptyset$. 
	For a minimal transversal $T \in \Tr(\cH)$,  the following statements are satisfied.
	\begin{description}
		\item[(i)] If $\eta(T)=\{0\}$, then $T\cap V_1 \in \cT_0$ and $T\cap V_2 \in \Tr(\cH_2)$. 
		\item[(ii)] If $\eta(T)=\{i\}$ for $i=1,2$, then $T\cap V_1 \in \cT_i$ and $T\cap V_2 \in \Tr(\cF_i')$. 
		\item[(iii)] If $\eta(T)=\{i,j\}$, then $T\cap V_2 \in \Tr(\cH_2)$, and moreover,   
		exactly one of the following two conditions holds. 
		
		{\bf (iii-1)} $T\cap V_1 \in \cT$.
		
		{\bf (iii-2)} There exists a vertex $v$ such that 
		either  $\bigl(T\cap S_i=\{v\}$ and $(T\setminus \{v\})\cap V_1 \in \cT_j\bigr)$ or $\bigl(T\cap S_j=\{v\}$ and $(T\setminus \{v\})\cap V_1 \in \cT_i\bigr)$. 
		%%%and $T\cap V_2$ is not a transversal of $\cF_j'$.  
		
		\item[(iv)] If $\eta(T)=\{0,1,2\}$, then  $T\cap V_1 \in \cT$ and $T\cap V_2 \in \Tr(\cH_2)$. 
	\end{description}
\end{lemma}

\noindent
\proof
Let $T$ be a minimal transversal of $\cH$. 
Since $\cP\not=\emptyset$, we have $\eta(T)\not=\emptyset$.

{\bf (i)} and {\bf (ii)}. Let $\eta(T)=\{i\}$ for some $i$. 
We first show that $T\cap V_1$ is a minimal transveral of $\cH_1$, implying that  
$T\cap V_1 \in \cT_i$. 
By definition, $T\cap V_1$ is a transveral of $\cH_1$. Since $T \in \Tr(\cH)$, for each $w \in T \cap V_1$, there exists a hyperedge $H_w \in \cH$ such that $H_w \cap T =\{w\}$. 
For $w  \in T \cap (V_1 \setminus S_i)$, we have 
$H_w \in \cH_1$, implying that $(T\setminus \{w\})\cap V_1$ is not a transveral of $\cH_1$. 
For $w \in T \cap S_i$, if  $T \cap S_i=\{w\}$ holds, then we have $(T\setminus w) \cap P=\emptyset$ for a $P \in \cP\,(\subseteq \cH_1)$.
Thus $T\cap V_1$ is a minimal transveral of $\cH_1$. 
On the other hand, if $|T \cap S_i|\geq 2$, we have $H_w \in \cH_1$, since any $H \in F_j$ ($j\not=i$) contains $S_i$. 
This implies that $T\cap V_1$ is a minimal transveral of $\Tr(\cH_1)$.

If $\eta(T)=\{0\}$, for any $F \in \cF_1 \cup \cF_2$, we have $F \cap T\not=\emptyset$.
This implies that  $T\cap V_2 \in \Tr(\cH_2)$.  

We next show that $T\cap V_2 \in \Tr(\cF_i')$ if $\eta(T)=\{i\}$ for $i=1,2$. 
Since $T\cap (\cup_{j\not=i}S_j)=\emptyset$, $T\cap V_2$ is a transversal of  $\cF_i$. 
Since $T$ is a minimal transversal of $\cH$, for each $w \in T\cap V_2$, there exists a hyperedge $H_w \in \cH$ such that $H_w \cap T=\{w\}$. Since $w \in V_2$ and $T \cap S_j=\emptyset$ for $j\,(\not=i)$, $H_w$ is contained in $\cF_i\cup \cH_2$.
Thus  $T\cap V_2$ is a minimal transversal of  $\cF_i \cup \cH_2$, which implies that  $T\cap V_2 \in \Tr(\cF_i')$. 

{\bf (iii)}. Let $\eta(T)=\{i,j\}$.
By definition, we have $T\cap V_2 \in \Tr(\cH_2)$.  
Let us assume that $T\cap V_1 \not\in \cT$, and show that  {\bf (iii-2)} holds. 
Since $T \in \Tr(\cH)$, for each $w \in T \cap V_1$, there exists a hyperedge $H_w \in \cH$ such that $H_w \cap T =\{w\}$. 
For $w  \in T \cap (V_1 \setminus (S_i\cup S_j))$, we have 
$H_w \in \cH_1$, implying that $(T\setminus \{w\})\cap V_1$ is not a transveral of $\cH_1$. 
However, 
since $T\cap V_1 \not\in \cT$, there exists a $v \in T\cap V_1$ such that $T\setminus \{v\}\cap V_1$ is a transversal of $\cH_1$. Thus $v$ is contained in $S_i \cup S_j$.  
Let us assume that $v \in S_i$, since otherwise we exchange $i$  with  $j$.  
For this $v$,  we have $H_v \in \cF_1 \cup \cF_2$.
This means that $j\not=0$ and $H_v \in \cF_j$. 
Since $S_i \subseteq H_v$, $T\cap S_i=\{v\}$ holds. 
%Moreover, $H_v \in \cF_j$ implies that $T\cap V_2$ is not a transversal of $\cF_j'$.  

We next show that $(T\setminus \{v\})\cap V_1 \in \cT_j$.
If $|T\cap S_j|=1$, say $T\cap S_j=\{u\}$ for some $u$, then  
$(T\setminus \{v,u\}) \cap V_1$ is not a transversal of $\cH_1$, since $\cP\not=\emptyset$. 
Thus in this case $(T\setminus \{v\})\cap V_1$ is a minimal transversal of $\cH_1$, and is contained  in $\cT_j$.
On the other hand, if $|T\cap S_j| \geq 2$, then we have $H_u \in \cH_1$ for any $u \in T\cap S_j$.
Thus  it holds that $(T\setminus \{v\})\cap V_1 \in \cT_j$.

{\bf (iv)}. By definition, we have $T\cap V_2 \in \Tr(\cH_2)$.   
Since $T \in \Tr(\cH)$, for each $w \in T \cap V_1$, there exists a hyperedge $H_w \in \cH$ such that $H_w \cap T =\{w\}$. 
For $w  \in T \cap (V_1 \setminus (S_0\cup S_1 \cup S_2))$, we have 
$H_w \in \cH_1$, implying that $(T\setminus \{w\})\cap V_1$ is not a transveral of $\cH_1$. 
For $w  \in T \cap (S_0\cup S_1 \cup S_2)$, we again have $H_w \in \cH_1$, implying that $(T\setminus \{w\})\cap V_1$ is not a transveral of $\cH_1$. 
Therefore, $T\cap V_1 \in \cT$ holds. 
\qed

By Lemma \ref{lemma-maxk12}, we can compute in polynomial time $\Tr(\cH)$ from $\Tr(\cH_1)\,(=\cT \cup \cT_0 \cup \cT_1 \cup \cT_2)$ and 
$\Tr(\cH_2)$, and $\Tr(\cF'_i)$ ($i=1,2)$ if $\cP\not=\emptyset$.

\begin{lemma}
	\label{lemma-maxk12edg}
	Let $\cH$ be a hypergraph that satisfies \raf{eq--aed1} and $\cP\not=\emptyset$. 
	Then we have 
	\[
	\Tr(\cH) \subseteq \Bigl(\cT_0 \cup \cT \cup \bigcup_{i=1,2} \{ T \cup \{v\} \mid T \in \cT_i, v \in S_0 \cup S_{3-i}  \}\Bigr) \wedge \Tr(\cH_2) \cup  \bigcup_{i=1,2} (\cT_i \wedge \Tr(\cF_i')).  
	\]
\end{lemma}

The following two lemmas show that  $|\Tr(\cH_1)|$, $|\Tr(\cH_2)|$, and $|\Tr(\cF'_i)|$ ($i=1,2)$ 
are bounded by the size of $\Tr(\cH)$, which immediately implies that  the sum of their volume is bounded by the volume of $\cH$.

\begin{lemma}
	\label{lemma-maxk13}
	Let $\cH$ be a hypergraph that satisfies \raf{eq--aed1}. 
	Then the following statements holds.
	\begin{description}
		\item[(i)] If $\cT_0\not=\emptyset$, then we have $\cT_0\dot{\wedge}\Tr(\cH_2) \in \Tr(\cH)$. 
		\item[(ii)] For $i=1,2$, we have $\cT_i\not=\emptyset$ and $\cT_i\dot{\wedge}\Tr(\cF_i') \in \Tr(\cH)$. 
		\item[(iii)] If $\cT \not=\emptyset$ and $\cP\not=\emptyset$, then $\cT \dot{\wedge}\Tr(\cH_2) \in \Tr(\cH)$. 
	\end{description}
\end{lemma}

\noindent
\proof
{\bf (i)}. For each $T_1 \in \cT_0$ and $T_2\in \Tr(\cH_2)$, $T_1 \cup T_2$ is a transversal of $\cF_j$ for any $j$. 
Thus $T_1 \cup T_2$ is a minimal transversal of $\cH$. 
{\bf (ii)}. For $i=1,2$, $\cT_i=\emptyset$ if and only if there exists a hyperedge $H$ such that $H \subseteq \cup_{j\not=i}S_j$. Since $\cH$ is Sperner, the latter implies that $\cF_i=\emptyset$. 
Thus $\cF_i\not=\emptyset$ implies $\cT_i\not=\emptyset$. 
Morever, for each $T_1 \in \cT_i$ and $T_2\in \Tr(\cF_i')$, $T_1 \cup T_2$ is a transversal of $\cF_j$ for $j \,(\not=i)$. 
Thus $T_1 \cup T_2$ is a minimal transversal of $\cH$. 

{\bf (iii)}. For each $T_1 \in \cT$ and $T_2\in \Tr(\cH_2)$, $T_1 \cup T_2$ is a transversal of $\cF_j$ for any $j$, 
since $\eta(T_1)\geq 2$. 
Thus $T_1 \cup T_2$ is a minimal transversal of $\cH$. 
\qed

As a corollary of Lemma \ref{lemma-maxk13}, we have the following result. 
\begin{lemma}
	\label{lemma-maxk14}
	Let $\cH$ be a hypergraph that satisfies \raf{eq--aed1}, $\cP\not=\emptyset$ and $\cT\not=\emptyset$. 
	Then we have 
	%\begin{description}
	%\item[(i)] If $\cF_i\not=\emptyset$ for all $i$, then 
	\begin{equation}
	(\cT \dot{\cup} \cT_0) \dot{\wedge}\Tr(\cH_2) \,\ \dot{\cup} \ \  \cT_1\dot{\wedge}\Tr(\cF_1')  \,\ \dot{\cup} \ \  \cT_2\dot{\wedge}\Tr(\cF_2')  
	\ \subseteq \ \Tr(\cH). 
	\end{equation}
	%\item[(ii)] If $\cF_j=\emptyset$ and $\cF_i\not=\emptyset$ for $i$ with $i\not=j$, then 
	%\begin{equation}
	%(\cT_o \,\dot{\cup} \,\cT_j)\dot{\wedge}\Tr(\cH_2) \,\ \dot{\cup} \, \ \ \dot{\bigcup_{i\not=j}}\cT_i\dot{\wedge}\Tr(\cF_i')  
	%\ \subseteq \ \Tr(\cH). 
	%\end{equation}
	%\end{description}
\end{lemma}

\noindent
\proof
Immediate from Lemma \ref{lemma-maxk13}. 
%{\bf (ii)}. Since $\cF_j=\emptyset$,  we have $\cT_j \dot{\wedge}\Tr(\cH_2) \subseteq \Tr(\cH)$. This together with 
%Lemma \ref{lemma-maxk13} implies the statement.
\qed

We remark that   $\cT$, $\Tr(\cH_2)$, $\cT_i$ $(i=1,2)$, and $\Tr(\cF_i')$ $(i=1,2)$ in  Lemma \ref{lemma-maxk13} are all nonemtpy. However,  $\cT_0$ might be empty.   

Based on Lemmas \ref{lemma-maxk12edg} and \ref{lemma-maxk14}, in {\bf Case III}, 
our procedure decomposes $\cH$ into 
$4$ hypergraphs $\cH_1$, $\cH_2$, and $\cF_i'$ $(i=1,2)$.

\medskip

%\noindent{\bf Analysis of the running time.}~ 
Let $n_1=|V_1|$, $n_2=|V_2|$, $m_1=|\cH_1|$, $k_1=|\Tr(\cH_1)|$, $m_2'=|\cH_2|$, $k_2=|\Tr(\cH_2)|$, $m_3=|\cF_1'|$, $k_3=|\Tr(\cF_1')|$,  $m_4=|\cF_2'|$ and $k_4=|\Tr(\cF_2')|$. Note that $n(\cH_1)=n_1$, $n(\cH_2)=n(\cF_1')=n(\cF_2')=n_2$. By the assumptions in this case, $n_1+n_2=n$, $n_1,n_2\ge 1$, and $m_1,m_3,m_4\ge1$.
Let $k_{10}=|\cT_0|+|\cT|$, $k_{11}=|\cT_1|$ and $k_{11}=|\cT_2|$. Then $k_1=k_{10}+k_{11}+k_{12}$  and by the assumptions of this case and Lemma~\ref{lemma-maxk13}, we have $k_{10},k_{11},k_{12},k_2,k_3,k_4\ge 1$. Moreover, Lemma~\ref{lemma-maxk14} implies that 
\begin{equation}
\label{max-e1}
k_{10}k_2+k_{11}k_3+k_{12}k_4\le k.
\end{equation}
Let $\mu_1=\mu(\cH_1)$, $\mu_1=\mu(\cH_2)$, $\mu_3=\mu(\cF_1')$ and $\mu_4=\mu(\cF_2')$.  
If $m_2'=0$ then $\Tr(\cH_2)=\{\emptyset\}$ and hence we do not recurse on $\cH_2$, giving thus the recurrence:
\begin{align}\label{rec1-1--}
T(\mu)\le 1+T(\mu_1)+T(\mu_3)+T(\mu_4).
\end{align}
In the general case, we get the recurrence
\begin{align}\label{rec1-1-}
T(\mu)\le 1+T(\mu_1)+T(\mu_2)+T(\mu_3)+T(\mu_4).
\end{align}
Note that \raf{max-e1} implies by Fact~\ref{fact2} that $\sum_{i=1}^4k_i\le k+3$. Using $n_1,n_2\ge 1$ and $k_1,k_2+k_3+k_4\ge 3$, we get by Fact~\ref{fact1} that 
\begin{align}\label{e-r}
\mu_1+\mu_2+\mu_3+\mu_4&=n_1m_1k_1+n_2m_2'k_2+n_2m_3k_3+n_2m_4k_4\nonumber\\
&\le m[n_1k_1+n_2(k_2+k_3+k_4)]\le m[3+(n-1)k]=\nonumber\\&m[nk-(k-3)]\le\mu-1,
\end{align}	
for $k\ge 4$. Using \raf{e-r} in \raf{rec1-1--} and \raf{rec1-1--}, we get by induction that $T(\mu)\le\mu$ in both cases.

\subsubsection*{Case IV}

By Lemma \ref{lemma-maxk12}, we can compute in polynomial time $\Tr(\cH)$ from $\Tr(\cH_1)$ and 
$\Tr(\cH_2)$, and $\Tr(\cF_i)$ ($i=1,2)$.
However, in Case {\bf IV}, $\cT\cup \cT_0$ might be empty. Hence, 
it is not clear from Lemma \ref{lemma-maxk14} that the sum of their volume $\sum_{i=1}^2(\mu(\cH_i)+\mu(\cF_i'))$ is  bounded by $\mu(\cH)$. 
We therefore deompose $\cH$ in the way different from {\bf Case III}, if $\cT_0$ is empty. 

Since $\cP\not=\emptyset, \{S_0\cup S_1 \cup S_2\}$, there exist a vertex $v^* \in S_0\cup S_1 \cup S_2$ and a
hyperedge $P \in \cP$ such that $v^* \not\in P$.
Let us assume without loss of generality that $v^* \in S_0 \cup S_2$, and  define
\begin{eqnarray}
\cF_1^*&=&\{F \cup \{v^*\} \mid F \in \cF_1^{V_2}\}\cup \cH_2. 
\end{eqnarray} 
In other word, any hyperedge $H^* \in \cF_1^* \setminus \cH_2$ is obtained from $H \in \cF_1$ by replacing $S_0 \cup S_2$ by vertex $v^*$.
Let $\cT^0$ be a family of minimal transversals $T$ of $\cF_1^*$ with $v^*\not\in T$, and 
let $\cT^1$ be a family of minimal transversals $T$ of $\cF_1^*$ with $v^*\in T$.  

\begin{lemma}
	\label{lemma-kz46-1}
	Let $\cH$ be a hypergraph that satisfies \raf{eq--aed1}. 
	Then
	%\begin{description}
	%\item[(i)]$(\cT^1)^{V_2} \subseteq \Tr(\cH_2)$. 
	%\item[(ii)]
	we have $\Tr(\cH_2) \subseteq (\cT^1)^{V_2} \cup \cT^0$ and $\cT^0= \Tr(\cF_1')$. 
	%\end{description}
\end{lemma}

By Lemmas \ref{lemma-maxk12} and \ref{lemma-kz46-1}, we have the following inclusion. 

\begin{lemma}
	\label{lemma-maxk12edg1}
	Let $\cH$ be a hypergraph that satisfies \raf{eq--aed1}. Assume that $\cT,\cT_0=\emptyset$ and $\cP\not=\emptyset$. 
	Then we have 
	\[
	\Tr(\cH) \subseteq \Bigl(\bigcup_{i=1,2} \{ T \cup \{v\} \mid T \in \cT_i, v \in S_0 \cup S_{3-i}  \}\Bigr) \wedge 
	\Bigl((\cT^1)^{V_2} \cup \cT^0\Bigr)
	\cup  \cT_1 \wedge \cT^0 \cup  \cT_2 \wedge \Tr(\cF_2').  
	\]
\end{lemma}
By this lemma, if $\cT,\cT_0=\emptyset$ and $\cP\not=\emptyset$,  
we can compute in polynomial time $\Tr(\cH)$ from $\Tr(\cH_1)\,(=\cT_1 \cup \cT_2)$ and 
$\Tr(\cF_1^*)\,(=\cT^0 \cup \cT^1)$, and $\Tr(\cF'_2)$ 

We next bound the size of three transversal hypergraphs $\Tr(\cH_1)$ and 
$\Tr(\cF_1^*)$, and $\Tr(\cF'_2)$.  
\begin{lemma}
	\label{lemma-kz46-2}
	Let $\cH$ be a hypergraph that satisfies \raf{eq--aed1} and $\cT=\emptyset$.
	Assume that $v^* \in S_0 \cup S_2$ and $P \in \cP$ satisfy $v^* \not\in P$. 
	%, and that 
	%$v^* \in S_1\cup S_2 \cup S_3$ and a
	%hyperedge $P \in \cP$ satisfy $v^* \not\in P$.
	Then for any $T \in \cT_1$ and any $U \in \cT^1$, 
	$T \cup U$ is a minimal transversal of $\cH$. 
\end{lemma}

\proof
By definition, $T \cup U$ is a transversal of $\cH$.
Since $U \in \cT^1$, for any $u \in U$,  $T \cup (U\setminus \{u\})$ is not a transversal of $\cH$.
Assuming that $T' \cup U$ is  a minimal transversal of $\cH$ such that $T' \subsetneq T$, we derive a contradiction. 
Since $T$ is a minimal transversal of $\cH_1$, $T'$ is not a  transversal of $\cH_1$.
However, by our assumption, $T' \cup \{v^*\}$ is a transversal of $\cH_1$, implying that there exist a minimal transversal $T'' \cup \{v^*\}$ of $\cH_1$ such that $T'' \subseteq T'$. 
Note that $T'' \cap S_1 \not=\emptyset$, since we have a hyperedge $P\,(\in \cP)$ with $v^*\not\in P$. 
This implies that $T'' \cup \{v^*\} \in \cT$, a contradiction. \qed

\begin{lemma}
	\label{lemma-maxk46-36}
	Let $\cH$ be a hypergraph that satisfies \raf{eq--aed1} and $\cT=\emptyset$.
	Let $v^* \in S_0 \cup S_2$ and $P \in \cP$ with $v^* \not\in P$. 
	Then we have 
	\begin{equation}
	\cT_1 \dot{\wedge} \Tr(\cF_1^*)\,\ \dot{\cup}    \,\ \cT_2\dot{\wedge}\Tr(\cF_2')  \ \subseteq \ \Tr(\cH). 
	\end{equation}
\end{lemma}

\noindent
\proof
By definition and Lemma \ref{lemma-kz46-1}, 
we have $\Tr(\cF_1^*)=\cT^0 \dot{\cup} \cT^1=\Tr(\cF_1')\dot{\cup} \cT^1$. 
Thus, 
\[\cT_1 \dot{\wedge} \Tr(\cF_1^*)\,\ \dot{\cup}    \,\ \cT_2\dot{\wedge}\Tr(\cF_2') =
\cT_1\dot{\wedge} \cT^1 \,\ \dot{\cup}    \,\ \cT_1\dot{\wedge}\Tr(\cF_1') \,\ \dot{\cup} \,\ \cT_2\dot{\wedge}\Tr(\cF_2')
\]
By Lemmas \ref{lemma-maxk13} and \ref{lemma-kz46-2}, it is contained in $\Tr(\cH)$.  
\qed

In {\bf Case IV}, 
if $\cT_0$ is nonempty, then we decompose $\cH$ in the way described in {\bf Case III}. 
Otherwise, 
based on Lemmas \ref{lemma-maxk12edg1} and \ref{lemma-maxk46-36},  
we decomposes $\cH$ into 
$3$ hypergraphs $\cH_1$, $\cF_1^*$,  and $\cF_2'$, 
where we assume that $v^* \in S_0 \cup S_2$. 
We note that $\Tr(\cH)$ can be computed from $\Tr(\cH_1)$, $\Tr(\cF_1^*)$,  and $\Tr(\cF_2')$
in $L_2(\mu)=\poly(n,m,k)$ time. 

In order to analyze $T(\mu)$, let 
$n_1=|V_1|$, $m_1= |\cH_1|$, $k_1=\Tr(\cH_1)$,    
$n'_2=|V_2|+1$, $m'_2= |\cF_1^*|$, $k_2=\Tr(\cF_1^*)$, 
$n_3=|V_2|$, $m_3= |\cF_2'|$, and $k_3=\Tr(\cF_2')$.  
Note that all are positive, $n_1,n'_2\leq n-1$, $n_1+n_2=n+1$, and $m_1,m_2',m_3'\leq m$. 
Since $|\cT_1|, |\cT_2|, k_2, k_3 \geq 1$, it follows from Fact \ref{fact2} and Lemma \ref{lemma-maxk46-36} that $k_1,k_2\leq k$ and  
$k_1+k_2+k_3\leq k+2$. Thus,
\begin{eqnarray}
n_1m_1k_1+n'_2m'_2k_2+n_3m_3k_3 &\leq& (n_1k_1+n'_2k_2+n_3k_3)m \nonumber\\
&\leq& (n_1k_1+n'_2(k_2+k_3))m \nonumber\\
&\leq& ((n-1)k+ 4)m \label{eq-89edes}\\
&\leq& nmk-1, \label{eq-89edes1}
\end{eqnarray}
where \raf{eq-89edes} is obtained from Fact \ref{fact1} and 
\raf{eq-89edes1} is obtained from $k\geq 5$. 
This implies that $T(\mu) \leq \mu$. 

\subsection{$3$-sum decomposition - case 3}

Let 
$\cH_1=\cH_{V_1}$ and $\cH_2=\cH_{V_2}$. 
By Corollary \ref{c-decomp}, we have two nonempty disjoint sets $S_1$ and $S_2$ in $V_2$.  
Define  
\begin{eqnarray}
\cF_0&=&\{H \in \cH \mid H \cap V_1= S_1 \cup S_2, H \cap V_2\not=\emptyset\},\\
\cF_1&=&\{H \in \cH \mid H \cap V_1=  S_2, H \cap V_2\not=\emptyset\},\\
\cF_2&=&\{H \in \cH \mid H \cap V_1= S_1, H \cap V_2\not=\emptyset\}. 
\end{eqnarray} 
Note that $V_1,V_2\not=\emptyset$, $\cH_1\not=\emptyset$, and at least two of $\cF_i$ are nonempy. 
$\cH$ can be partitioned in the following way.
\begin{equation}
\label{eq--aed1-a}
\cH=\cH_1 \dot{\cup} \cH_2\dot{\cup} \cF_0  \dot{\cup} \cF_1 \dot{\cup} \cF_2, 
\end{equation}
where $\dot{\cup}$ denotes the disjoint union. 
For $i=1,2$, let 
\begin{equation}
\cT_i=\{T \in \Tr(\cH_1) \mid T\cap S_i \not=\emptyset, T \cap S_j=\emptyset\,\, (j\not=i)\},
\end{equation} 
and let 
\begin{equation}
\cT=\Tr(\cH_1) \setminus (\cT_1 \cup \cT_2).
\end{equation}
%By the definition, $\cT_0$ consists of minimal transversals of $\cH_1$ that intersect at least two of $S_i$, and  
By definition, we have 
\begin{equation}
\Tr(\cH_1)=\cT  \dot{\cup} \cT_1 \dot{\cup} \cT_2.  
\end{equation}
Let 
\begin{equation}
\cP=\cH_{ S_1\cup S_2}\,\,(=\{H \in \cH \mid H \subseteq S_1 \cup S_2\}). 
\end{equation}

We separately consider the following 4 cases, similar to the case analysis in Section 
\ref{section-3-sum-case2}

\medskip
\noindent
{\bf Case I}: $\cP=\emptyset$. 

\smallskip

\noindent
{\bf Case II}: $\cP=\{S_1 \cup S_2\}$. 

\smallskip

\noindent
{\bf Case III}: $\cP\not= \emptyset, \{S_1 \cup S_2\}$ and  $\cT\not=\emptyset$. 

\smallskip

\noindent
{\bf Case IV}: $\cP\not= \emptyset, \{S_1 \cup S_2\}$ and  $\cT=\emptyset$. 

\medskip
\noindent

In {\bf Case I}, we  can decompose $\cH$ into   $\cH_{1}$ and $\cH_{S_1\cup S_2\cup V_2}$, where the proof is similar to {\bf Case I} in Section \ref{section-3-sum-case2}. 
In {\bf Case II}, we note that $|\cH_1|\geq 2$ holds. 
To see this, recall from Corollary \ref{c-decomp} that $\cH$ is irredundant Sperner and 
has neither identical vertices nor vertices of degree $1$. 
Moreover, we have $|\cH_1|+|V_1|\geq 4$ from $3$-sum--case 3 (ii) in Corollary \ref{c-decomp}, which implies  $|\cH_1|\geq 2$. 
We therefore decompose $\cH$ into   $\cH_{1}$ and $\cH_{S_1\cup S_2\cup V_2}$ in a way similar to  {\bf Case II-I} in Section \ref{section-3-sum-case2}, where the proof is also similar to {\bf Case II-I} in Section \ref{section-3-sum-case2}. 
For the other cases, we note that $\cF_0=\emptyset$ since $\cP\not=\emptyset$ and $\cH$ is Sperner.
Therefore, decompositions similar to the corresponding cases in Section \ref{section-3-sum-case2} work properly.

%\section{Dualizing network hypergraphs} 

\section*{Acknowledgements} 
We thank Endre Boros and Vladimir Gurvich for helpful discussions.
The research of the second author is supported by Grant-in-Aid for Scientific Research 24106002, 25280004, 26280001 and JST CREST Grant Number JPMJCR1402, Japan. 
\newcommand{\etalchar}[1]{$^{#1}$}

\appendix

\end{document}